\documentclass[journal ]{new-aiaa}
\usepackage[utf8]{inputenc}
\usepackage{textcomp}

\usepackage{graphicx}
\usepackage{amsmath}
\usepackage[version=4]{mhchem}
\usepackage{siunitx}
\usepackage{longtable,tabularx}
\usepackage{lineno}
\usepackage[normalem]{ulem}
\usepackage{caption}
\usepackage{subcaption}

\title{A Probabilistic Modeling Framework for Transient Debris Outcomes in Low Lunar Orbits}

\author{Arjun Chhabra\footnote{PhD Candidate, Mechanical and Aerospace Engineering, Princeton University, Princeton, NJ, 08544 U.S.A.} and Ryne Beeson\footnote{Assistant Professor, Mechanical and Aerospace Engineering, Princeton University, Princeton, NJ, 08544 U.S.A., AIAA Senior Member (Corresponding author, \href{mailto:ryne@princeton.edu}{ryne@princeton.edu})}}
\affil{Princeton University, Princeton, New Jersey, 08544, U.S.A.}

\begin{document}

\maketitle

\begin{abstract}
This work presents a novel probabilistic framework for the assessment of orbital debris fragment outcomes in the low lunar orbit (LLO) regime over time horizons ranging from a few hours to a hundred days after a debris generation event. The framework develops continuous distributions for the probability of sinking (lunar collision) and non-sinking over these transient time horizons, building upon NASA's Standard Breakup Model to provide insights into the variations in the likelihood of debris outcomes across the LLO regime and offering an alternative to computationally expensive Monte Carlo simulations. The effects of perturbative forces such as solar radiation pressure are used to assess the realization of debris outcomes over varying time horizons, providing an analytical framework that bounds the likelihood of each outcome. Results indicate variations in the probability of sinking over short time horizons depending on the originating location of the debris generation event, as well as a link between the physical characteristics of fragments and their likelihood of sinking over longer time horizons. The future incorporation of this framework into broader-scale orbital environment models, mission risk assessment procedures, and policy development are briefly discussed.
\end{abstract}

\setcounter{table}{0}

\section{Introduction}
\label{sec: introduction}

Over the course of the past decade, there has been a renewed international interest in returning humanity to the lunar surface and a corresponding increase in planned missions to key lunar orbits \cite{johnson_fly_2022, baker2024comprehensive, bukley2024moonstruck}. The onset of efforts such as the Artemis program and the International Lunar Research Station project indicate the developing foundations for a wide-spanning set of international endeavors towards establishing a sustained presence in the lunar and cislunar (i.e., Earth-Moon) domain. Notably, these missions are increasingly supported by involvement of commercial space actors (such as through NASA's Commercial Lunar Payload Services program), with potential prospects for entirely independent missions adding to a growing population of resident space objects (RSOs) in key orbits. With this foreseeable growth in the volume of upcoming missions, as well as an evolution in the complexity of their operations \cite{vedda_cislunar_2018, ChhabraJSPCEOs}, it becomes critical to ensure that desirable orbital resources in the cislunar domain are sustainably used and remain viable destinations for missions in the long-term. A key factor in such considerations of long-term orbital sustainability is the minimization of risks posed by orbital debris generation in lunar and cislunar orbits, such that individual missions may maximize their likelihood of success and the orbital environment as a common-pool resource \cite{BWOstrom} remains fully operational in the coming decades.

Orbital debris generation (whether through the release of ejecta, on-orbit breakup events, or collisions between RSOs) leads to the presence of free-drifting debris fragments in the orbital environment. These fragments may pose risks to the success of active missions through potential conjunctions, which in turn could generate additional debris; they could additionally linger indefinitely in the orbital environment, potentially rendering certain orbital regions inoperable for future missions. Since Kessler's initial identification \cite{kessler_collision_1978} of debris generation as a significant risk to the long-term viability of orbital resources, significant strides have been made in characterizing the trajectories of generated debris fragments and developing models to forecast the evolution of the debris population in the orbital environment. It is important to note, however, that these efforts have predominantly been focused on near-Earth orbital regimes, owing to the majority of space activity being concentrated in these regimes since the Apollo era. To begin, these efforts focused on using ground-based observations of RSO breakup events for the development of semi-empirical probabilistic models (such as NASA's EVOLVE4.0 \cite{johnson_nasas_2001}) for individual debris generation events and the assignment of physical characteristics for the resulting debris fragments. Subsequently, efforts were focused on developing larger-scale models of debris fluxes and impact directions (e.g., NASA's Orbital Debris Engineering Models (ORDEM) \cite{krisko2014new, matney2023overview} and ESA's MASTER \cite{klinkrad1997introduction, Horstmann2020Master} model), informing the long-term evolutionary modeling and forecasting of the near-Earth debris population (e.g., via NASA's LEGEND model \cite{liou2004legend}). Insights from the use of such models across have further led to the development of technical standards\cite{national_aeronautics_and_space_administration_process_2021, ISO}, guidelines \cite{UN_LongTerm_2019}, and policies (such as the U.S. Government's \textit{Orbital Debris Mitigation Standard Practices} \cite{ODMSP}) towards the mitigation of orbital debris generation and associated risks for near-Earth missions.

Additional modeling efforts in the near-Earth case have also sought to enable forecasting of the debris population for future scenarios of the orbital environment, under various assumptions regarding launch rates, the emergence of high-density regions, or certain mitigation strategies being enacted. Such efforts reflect a rapidly growing RSO population in LEO -- owing partially to the increasing involvement of commercial actors and the onset of large multi-satellite constellations \cite{schweiger2023survey} in recent years -- and the associated uncertainty in the realized growth of the RSO population in these orbital regimes over the coming decades. The time evolution of the initial RSO population (as parameterized by the various assumptions listed above) may be modeled via propagation using Monte Carlo techniques \cite{dambrosio_capacity_2022}, or through source-sink evolutionary models based on altitude shells \cite{jang_jang_2023} and other continuum-based density models \cite{GIUDICI2024115} which capture the aggregate projected population growth over time. Forecasts from these models can then be used to assess orbital carrying capacity \cite{dambrosio_capacity_2022}, space traffic management needs \cite{gregoire2023validation}, or the applicability of new policies \cite{borowitz2024exploring} across a gamut of potential growth pathways.  Several of these models have additionally been used to validate standardized reference scenarios of projected population growth across the orbital environment \cite{lifson2024development, lifson2025initial} and informing discussions of associated variations in the sustainability of near-Earth orbital resources. 

As the RSO population in lunar and cislunar orbits begins to grow, analogous development of models and associated analyses will likewise be critical to mitigate the potential degradation of key orbital resources and inform new standards, guidelines, and policies for cislunar missions \cite{gangestaad_pack_2023, ChhabraJSP2025}. While some efforts \cite{pasiecznik2023evolutionary} towards extending the capability of near-Earth models to cislunar orbits have been undertaken, the complexity of the dynamics which govern the motion of RSOs in various orbital regions of the Earth-Moon system will likely necessitate the development of new forecasting and risk assessment models specific to the cislunar domain. Space objects in lunar and cislunar orbits experience the gravitational effects of both the Earth and the Moon, which lead to highly nonlinear and chaotic free-drift dynamics within the cislunar domain \cite{KLMR2022}. As a result, the trajectories of debris fragments are quite sensitive to dynamical perturbations, and fragments generated in one orbital location with sufficient total mechanical energy could rapidly proliferate towards other spatially distant orbital regions. Additionally, mechanisms for fragments to be removed from the orbital environment (e.g., through lunar impact or heliocentric escape) are not as-yet sufficiently characterized in a domain-wide sense, and may carry associated risks in terms of removed fragments re-entering the cislunar domain after the end of the mission lifetime. Consequently, several modeling assumptions such as partitions in the position space or source-sink frameworks may not fully capture the complexity of cislunar debris risks in an aggregate sense; these challenges are similarly reflected in the predominant focus of existing literature towards studying cislunar debris generation events in narrower, mission-specific Monte Carlo-based analyses. For example, preliminary studies for the proposed Lunar Gateway characterized debris risks for its nominal orbit \cite{zimovan_2017_near} in terms of the release of ejecta  \cite{davis_disposal_2019}, post-mission disposal \cite{davis_disposal_2019, boudad_disposal_2018, davis_heliocentric_2019, davis_lunar_2021, guardabasso_lunar_2021, franz2025iac, george2024low}, clear-out times \cite{guardabasso_analysis_2023}, loitering \cite{scheuerle2025relative}, and recontact with the nominal orbit \cite{Chhabra_IACc} among others. Similar approaches have also been applied towards considering hypothetical debris generation events between existing missions \cite{gangestad2025iac}, collision risks from generated debris fragments to existing assets \cite{tonnini2025collision}, modeling uncertainty growth for conjunction assessments \cite{franzcislunar}, characterizing requisite keep-out zones \cite{Chhabra2025KeepOut}, or simulating debris generation events on reference trajectories from past missions and stable libration points of the Earth-Moon system \cite{boone_cislunar_2021}. The results from these efforts collectively reflect the sensitivity of fragment trajectories to the originating location of the debris generation event as well as to dynamical perturbations, with fragments often rapidly departing the originating orbit and proliferating through the accessible regions of the domain. 

To gain a broader sense of the aggregate evolution of the debris population, research efforts have also investigated the behavior of debris fragments generated within a subset of the cislunar domain by choosing to focus on a specific family of orbits. Notably, Low Lunar Orbits (LLOs) are a highly desirable family of orbits for missions in the immediate future \cite{johnson_fly_2022}, given their utility towards monitoring and characterizing the lunar surface or providing support to surface activities as part of orbiter-lander missions. Investigations of debris behavior in LLOs have the added benefit of the underlying dynamics being a close analogue to LEOs in the near-Earth case, with lunar gravitational effects being the dominant force (i.e., validity of a two-body dynamical model) and the primary perturbative effects arising from the irregularities in the lunar gravity field. Work by Boone et al. has explored the outcomes of debris fragments generated in key orbits in the LLO regime \cite{boone2023simulation} as well as for a broader set of randomly sampled originating locations throughout the LLO family \cite{boone2024monte, boone2025simulation}. Through their Monte Carlo simulations, their work studies the fraction of generated fragments which linger in the vicinity of the originating orbit over a period of one year, and characterizes the probability of collision for a set of polar and equatorial spacecraft with the generated fragments. They find that while collision probabilities remain low and a significant fraction of generated fragments depart the vicinity of the originating orbit, the remainder of fragments linger around the originating orbit even at the end of the one-year period. Additionally, their work identifies the perilune distance of the originating location at which debris fragments are generated to be the most significant contributor to the fraction of fragments departing the vicinity of of the considered orbit, indicating that a majority of fragments in sufficiently low-altitude LLOs may eventually decay to the lunar surface. 

This work presents a novel contribution to the existing literature on the modeling of debris outcomes in LLOs through the development of a \textbf{probabilistic modeling framework for debris outcomes} over short (hours to days) and medium (days to months) time horizons, to assist mission designers and inform future development of larger-scale forecasting and flux models. We leverage NASA's EVOLVE4.0 Standard Breakup Model (SBM) \cite{johnson_nasas_2001} to develop a measure-theoretic framework within which the likelihood of debris outcomes (sinking to the lunar surface or lingering within the LLO family) may be cohesively assessed for a given originating location, providing probability distributions for each outcome and characterizing perturbative effects from sources such as solar radiation pressure over longer time horizons.

The necessity of modeling of debris behavior over these \textit{transient} time horizons is motivated by the results from the simulations conducted by Boone et al. \cite{boone2023simulation} showing up to 30\% of generated fragments rapidly departing the vicinity of the originating orbit within the first few days. While some of these fragments may be removed from the orbital environment soon after the debris generation event via lunar impacts (i.e., `sinking'), the behavior of both sinking and non-sinking fragments may pose immediate risks to proximate missions and must be rigorously modeled to inform the development of flux models analogous to ORDEM as well as any necessary conjunction avoidance techniques \cite{davis2022debris}. The proposed approach additionally provides the benefit of being based on distributions spanning the probability space of possible fragments that could be generated by the SBM at a given location, thereby mitigating the drawbacks of Monte Carlo-based analyses requiring the propagation of a large number of fragments for each considered debris generation event in order to obtain statistically significant results. 

The proposed framework consists of two components: the first focused on short time horizons to characterize the probability of fragments sinking within a fragment's first orbital period (typically on the order of hours), and the second focused on medium time horizons which study the effects of perturbative forces on the trajectories of the remaining non-sinking fragments. This two-pronged approach leverages the underlying dynamics governing the motion of RSOs in the LLO regime, since we may use the distribution of fragment $\mathbf{\Delta v}$ values \textit{at the time of generation} to obtain the distribution of their perilune distances, which in turn indicates the probability of generated fragments sinking within one fragment period without the need for any propagation. The remainder of fragments, which linger in the orbital environment over longer time horizons, are then addressed by the second component of the framework in the context of the effects of perturbations which may lead to them impacting the lunar surface over time. This second component similarly builds on results from Boone et al. \cite{boone2023simulation} which found that the empirical fraction of lingering fragments reduced over longer time horizons, with specific percentages of such reductions varying across the considered orbits. We employ a first-order averaging approach using the Milankovitch orbital element vectors to derive analytical expressions for the time-evolution of the eccentricity vector for the non-sinking fragments under the perturbations due to SRP, in a manner similar to the work of Rosengren et al. in the near-Earth case \cite{rosengren2013long}. This approach makes explicit the dependence of such perturbative effects on the physical characteristics of debris fragments; our framework thus provides a Bayesian posterior for the distribution of area-to-mass ratios for the non-sinking fragments, which is used in the initial sampling of the Milankovitch vectors. Collectively, these two components provide a comprehensive probabilistic framework for the consideration of debris outcomes over short and medium time horizons.

The following sections detail the development of our proposed framework, demonstrate the validation of results against a Monte Carlo baseline for both components, and discuss the utility of this framework for mission designers and policymakers alike. We begin by providing a brief theoretical background in Section \ref{sec: theoretical_background}, discussing the mechanics of the SBM, the dynamics of the LLO regime, and the formulation of the Milankovitch orbital element vectors. Section \ref{sec: modeling_sinking} provides the theoretical development of the short time horizon component of the framework, including discussions towards the fitting of relevant distributions and comparisons against a Monte Carlo baseline. Likewise, Section \ref{sec: modeling_nonsinking} presents the medium time horizon component, including an assessment of the relative magnitude of considered perturbative forces and the utilization of the Milankovitch vectors in assessing the outcomes of non-sinking fragments. Lastly, Section \ref{sec: results_utilization} provides an example of the potential utilization of the proposed framework for mission design via debris risk considerations during preliminary orbit selection and the formulation of suitable probabilistic risk thresholds for policy development, as well as identifying key follow-on directions for the incorporation of this framework in broader-scale environmental flux and forecasting models.  

\section{Theoretical Background}\label{sec: theoretical_background}

\subsection{Standard Breakup Model}\label{sec: debris gen}

When an RSO experiences a debris generation event, whether through a collision or an accidental on-orbit breakup, the resulting fragments must be assigned both physical and dynamical characteristics. Multiple models have been developed for this purpose, including NASA's EVOLVE4.0 SBM \cite{johnson_nasas_2001} which is utilized in this work. The SBM is a semi-empirical probabilistic model which uses ground radar data of debris generation events in near-Earth orbit to propose probability distributions from which the requisite characteristics for the generated fragments may be sampled. For the development of the probabilistic framework, we consider accidental on-orbit breakup events (e.g., due to improper passivation before disposal or other mission-critical subsystem failures during the mission lifetime) and the resulting fragments that are generated by the SBM for a given originating RSO. The model uses the mass of the originating RSO, $m_O$ (in kg), as well as a minimum characteristic length, $L_{c, \min}$ (in meters) as input parameters, and samples fragments in a manner that is agnostic to the originating location at which the debris generation event occurs. To begin, the model uses $L_{c, \min}$ to identify the number of fragments to be generated as: 
\begin{equation}
\label{eq:num_frag_SBM}
    N_f(L_{c, \min}) = S \cdot 6(L_{c, \min})^{-1.6},
\end{equation}
where $S$ is a type-dependent (i.e., whether the event is a collision or accidental on-orbit breakups) unitless number. For accidental on-orbit breakups, $S = 1$. The characteristic length, $L_c$, which serves as the independent variable for each fragment, is sampled next and is subsequently used as a parameter to sample other fragment characteristics. The model calculates a maximum characteristic length, $L_{c, \max}$, to define a range over which the fragment $L_c$ values are sampled: 
\begin{equation}
    L_{c, \max} = \left(\frac{6m_O}{92.937\pi}\right)^{2.26^{-1}},
\end{equation}
assuming the RSO to be a perfect sphere with the density of aluminum. The fragment $L_c$ values are then sampled using a power-law distribution:
\begin{equation}
     L_c = \left(y(L_{c, \max})^{p + 1} + (1- y)(L_{c, \min})^{p + 1}\right)^{\frac{1}{p + 1}},   \quad y \sim \text{Uniform}(0, 1) ,
\end{equation}
where $p = -2.6$ for accidental on-orbit breakups. Once the characteristic length for each fragment has been assigned, it can be used as a parameter to sample the fragment's area-to-mass ratio $\psi$ (in m$^{2}$ kg$^{-1}$) using distributions provided by Johnson et al. \cite{johnson_nasas_2001}, with distributions describing the assignment of $\psi$ for $L_{c} \geq 11$ cm and $L_{c} \leq 8$ cm and a corresponding bridging function for intermediate values of $L_{c}$. Denoting the collective set of these distributions as $D^{\psi}$, we can represent this process as: 
\begin{equation}
    \psi \sim D^{\psi}(\lambda_{c}),\quad     \lambda_{c} = \log_{10}(L_{c})
\end{equation}
Having obtained the area-to-mass ratios for each fragment, the fragment area $A$ (in m$^2$) and mass $m$ (in kg) can then be calculated as: 
\begin{equation}
    A = \begin{cases}
        0.540424 \cdot L_{c}^2 & \text{if } L_{c} < 0.167 \text{ cm} \\
        0.556945 \cdot L_{c}^{2.0047077} & \text{otherwise}
    \end{cases} , \quad
    m = \frac{A}{\psi}
\end{equation}
Note that since the total mass of the generated fragments must be equivalent to the originating mass $m_O$, the last 2-8 fragments are separately sampled from the last bin of the $L_c$ power-law distribution until the required total mass has been achieved \cite{Krisko2011Model}.

\begin{figure}[h!]
    \centering
    \includegraphics[width=0.95\linewidth]{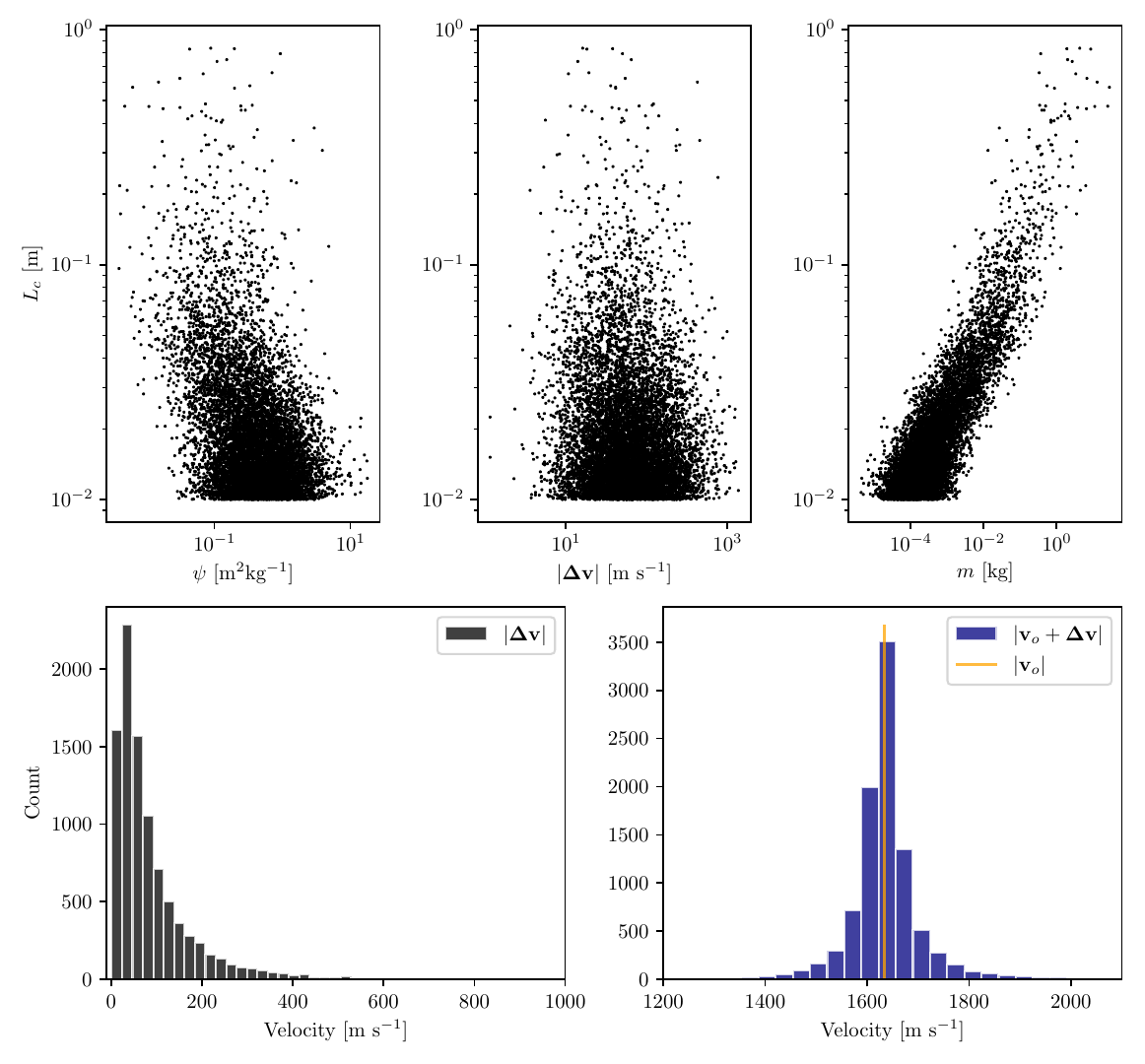}
    \caption{(top) Debris characteristics for one generation event. (bottom) Histograms of $|\Delta \text{v}|$ and net velocity magnitude.}
    \label{fig:debris_char}
\end{figure}

The final characteristic that needs to be assigned to each fragment is the fragment $\mathbf{\Delta v}$ (in m s$^{-1}$), which is assigned in two steps, with the magnitude being sampled based on $\psi$ as:  
\begin{equation}
\label{eq:dv_from_psi}
\begin{split}
    |\mathbf{\Delta v}|  &= 10^{y}, \quad   y \sim \mathcal{N}(\mu_{\mathbf{\Delta v}} =  0.2 \log_{10}(\psi) + 1.85, \sigma_{\mathbf{\Delta v}} = 0.4),
\end{split}
\end{equation}
and the direction being sampled uniformly on a unit sphere. To ensure that there is no clustering of samples near the poles of the unit sphere, directions are sampled using an inverse transform sampling approach \cite{marsaglia1972choosing, muller1959note}. Once a $|\mathbf{\Delta v}|$ value has been sampled, we may recover the Cartesian components of $\mathbf{\Delta v}$ as: 
\begin{equation}\label{eq:marginals_dv}
\begin{split}
     u \sim \text{Uniform}(-1, 1), &\quad \theta \sim \text{Uniform} (0, 2\pi)   \\
     \mathbf{\Delta v}_x = |\mathbf{\Delta v}|\sqrt{1-u^2}\cos \theta, \quad \mathbf{\Delta v}_y = |\mathbf{\Delta v}&|\sqrt{1-u^2} \sin\theta, \quad \mathbf{\Delta v}_z = |\mathbf{\Delta v}|u,
\end{split}
\end{equation}
where $u$ and $\theta$ represent the azimuthal and zenith components of the direction sampled via the inverse transform sampling approach. Once the $\mathbf{\Delta v}$ has been assigned, we may use the state of the originating RSO at the time of the debris generation event, denoted as $(\mathbf{r}_o, \mathbf{v}_o)$, to obtain the fragment's state as $  (\mathbf{r}, \mathbf{v}) = (\mathbf{r}_o, \mathbf{v}_o + \mathbf{\Delta v})$. Figure \ref{fig:debris_char} shows the values of these fragment characteristics -- specifically $m, \psi, \text{and } \mathbf{\Delta v}$ -- and their relationship to the independent variable, $L_{c}$, for one accidental on-orbit breakup for an RSO in a 100 km altitude polar circular LLO with mass $m_O=48.67$ kg\footnote{The choice of RSO mass here is to set $L_{c, \max} = 1.0$ m and provide an intuitive range for $L_{c}$.} and the model's $L_{c, \min} = 0.01$ m. Figure \ref{fig:debris_char} additionally shows the histogram of the magnitudes of fragment $\mathbf{\Delta v}$ and net velocities, with the latter showing a significant spread in the range on the order of $10^3$ m s$^{-1}$ centered around  the velocity magnitude of the originating location.

\subsection{Dynamics}
\label{sec: dynamics}

The free-drift motion of RSOs in the LLO family is governed primarily by the gravitational influence of the Moon, with the effects of additional celestial bodies and forces acting as perturbations. To model these dynamics at a sufficiently high fidelity, the effects of aspherical lunar gravity, third-body gravitational effects from the Sun and the Earth, and the acceleration due to solar radiation pressure (SRP) are considered. The free-drift motion of an RSO with a time-varying state $(\mathbf{r}, \mathbf{v}) \in \mathbb{R}^6$ defined in a Moon-Centered Inertial (MCI) reference frame\footnote{The MCI reference frame can be defined with the $\hat{i}$ direction along the intersection of the International Cartesian Reference Frame \cite{hughes2016general} and the Moon's equator at the J2000 epoch, the $\hat{k}$ direction along the Moon's spin axis at the J2000 epoch, and $\hat{j} = \hat{k} \times \hat{i}$.} and mass $m$, is described by the differential equation: 
\begin{equation}\label{eq: dynamics_overall}
    \ddot{\mathbf{r}} = -\frac{\mu}{|\mathbf{r}|^3}\mathbf{r} + \ddot{\mathbf{r}}_{\text{Asph. Gravity}} + \ddot{\mathbf{r}}_{\text{Third Body}} + \ddot{\mathbf{r}}_{\text{SRP}}\, ,
\end{equation}
where the first term is the two-body gravitational effect from the Moon, with $\mu = G(m + m_{Moon}) \approx Gm_{Moon}$ the gravitational parameter for the same. Each of the remaining terms in Eq. \eqref{eq: dynamics_overall} corresponds to the perturbative sources considered. To begin, we consider the aspherical nature of lunar gravity, since the mass distribution across the Moon is non-uniform; to incorporate the variations in the lunar gravity field due to the concentration of mass (`mascons') across the lunar surface \cite{muller1968mascons}, we utilize the Gravity Recovery And Interior Laboratory (GRAIL) Real-Time Gravity Model (GRGM) spherical harmonic lunar gravity model which was developed based on data from the GRAIL mission \cite{zuber2013gravity} and captures the effects of lunar mascons \cite{melosh2013origin}. These spherical harmonics may be written as: 
\begin{equation}\label{eq: spherical_harmonics}
    \begin{split}
        \ddot{\mathbf{r}}_{\textrm{Asph. Gravity}} &= \nabla_{\mathbf{r}} \, \mathcal{R}_{\textrm{GRGM}}(\mathbf{r} ; L) \\
        \mathcal{R}_{\textrm{GRGM}}(\mathbf{r} ; L) &:=  \mathcal{R}_{\textrm{GRGM}}(|\mathbf{r}|, \phi, \lambda ; L)   = \frac{\mu}{|\mathbf{r}|} \sum_{l = 2}^L \left(\frac{r_M}{|\mathbf{r}|}\right)^l \sum_{m = 0}^{l} P_{lm}(\sin{\phi})[C_{lm} \cos(m \lambda) + S_{lm}\sin(m \lambda)]\, ,
    \end{split}
\end{equation}
where $\mathcal{R}_{\textrm{GRGM}}(\mathbf{r} ; L)$ is the spherical harmonic gravitational potential defined up to degree and order $L$, using the RSO's position $\mathbf{r}$ to calculate the radius $|\mathbf{r}|$, latitude $\phi$, and longitude $\lambda$. The Legendre polynomials of degree $l$ and order $m$ are denotes as $P_{lm}$, and the GRGM provides the spherical harmonic coefficients $C_{lm}$ and $S_{lm}$ \cite{zuber2013gravity}. In a similar manner, the third-body gravitational effects from the Sun ($\odot$) and the Earth ($\oplus$) may be modeled as perturbing acceleration potential for each body \cite{scheeres2016orbital}:
\begin{equation}
        \ddot{\mathbf{r}}_{\textrm{Third Body}} = \nabla_{\mathbf{r}} \, \left(\mathcal{R}_{\oplus}(\mathbf{r}) + \mathcal{R}_{\odot}(\mathbf{r})\right), \quad
        \mathcal{R}_{\textrm{body}}(\mathbf{r}) = \mu_{\textrm{body}}\left( \frac{1}{|\mathbf{r} - \mathbf{d}_{\textrm{body}}|} - \frac{\mathbf{d}_{\textrm{body}} \cdot \mathbf{r}}{|\mathbf{d}_{\textrm{body}}|^3} \right) \, ,
\end{equation}
where $\mathbf{r}$ is the vector from the Moon to the RSO, $\mathbf{d}_{\textrm{body}}$ is the vector from the Moon to the perturbing body (in our case, the Earth or the Sun), and $\mu_{\textrm{body}} \approx Gm_{\textrm{body}}$ the gravitational parameter for the perturbing body. Lastly, the acceleration due to SRP is defined using a perturbing acceleration potential \cite{rosengren2013long, scheeres2016orbital}: 
\begin{equation}\label{eq: srp_def}
    \ddot{\mathbf{r}}_{\textrm{SRP}} = \nabla_{\mathbf{r}}\, \mathcal{R}_{\textrm{SRP}}(\mathbf{r}) , \quad \mathcal{R}_{\textrm{SRP}}(\mathbf{r}) = (1-\nu) p_{\textrm{SRP}} \, C_\textrm{R} \, \psi (\mathbf{r} \cdot \hat{\mathbf{r}}_{Sun}) \, ,
\end{equation}
where $p_{\textrm{SRP}} = 4.56 \cdot 10^{-6}$ N m$^{-2}$ is the solar pressure force, $C_\textrm{R}$ is the coefficient of reflectivity, $\psi$ is the area-to-mass ratio of the RSO, $\hat{\mathbf{r}}_{Sun}$ is the unit vector from the Sun to the RSO, and $\nu \in [0, 1]$ is the shadow function with zero corresponding to sunlit conditions and one corresponding to complete eclipse. The dynamics specified in Eq. \eqref{eq: dynamics_overall}, including all perturbations, are implemented in NASA's Generalized Mission Analysis Tool (GMAT) \cite{jah2009general, hughes2016general} using the associated Python API \cite{conway2020using}, and can be integrated with respect to time to obtain the time-evolution of the trajectory or orbit of a given RSO, $(\mathbf{r}_\tau, \mathbf{v}_\tau)_{\tau \in [0, t]}$ for any desired time horizon $t$. While the contributions of all perturbative forces are significant across the LLO regime, their relative magnitude (and thus their effect on the RSO's trajectory) in comparison to the two-body acceleration varies depending on the state of the RSO being considered. As discussed further in Section \ref{sec: dyn_rel_mag}, certain perturbations become more salient to the investigation of lingering fragments over longer time horizons. 

\subsection{Milankovitch Orbital Element Vectors}\label{sec: milankovitch_background}

While the definition of the RSO state in the Cartesian representation is useful to assess the time-evolution of its trajectory, other representations can additionally be used to more directly investigate the effects of various perturbations. In particular, this work utilizes the Milankovitch orbital element vectors \cite{milankovitch1941kanon}, which reformulate the problem of perturbation analysis using the angular momentum and eccentricity vectors from the unperturbed two-body problem. These vectors are specifically meaningful in enabling the assessment of first-order secular variations in an averaged sense \cite{allan1964long}, such that the long-term evolution of RSO orbits and trajectories due to a specific perturbative source may be evaluated. 

To begin, we define a Moon-Centered reference frame, aligned with the Earth-Sun Ecliptic plane as defined at the J2000 epoch. This reference frame (henceforth the MCE frame), has its $\hat{i}_{\textrm{MCE}}$ direction aligned with the Vernal Equinox, its $\hat{k}_{\textrm{MCE}}$ direction perpendicular to the Ecliptic plane at J2000, and the $\hat{j}_{\textrm{MCE}} = \hat{k}_{\textrm{MCE}} \times \hat{i}_{\textrm{MCE}}$ direction completing the right-handed frame. While the Milankovitch orbital element vectors may be defined in the previously mentioned MCI reference frame as well, the MCE reference frame has the additional benefit of the `apparent' Sun vector to the Moon remaining within the $\hat{i}_{\textrm{MCE}}-\hat{j}_{\textrm{MCE}}$ plane at all times. This vector, $\hat{\mathbf{r}}_\odot$, is defined as: 
\begin{equation}
\renewcommand{\arraystretch}{0.30}
    \hat{\mathbf{r}}_{\odot}(t ; \lambda) = \begin{pmatrix}
        \cos(\omega_{\odot} t + \lambda), & \sin(\omega_\odot t + \lambda), &0 
    \end{pmatrix}^T,
\end{equation}
where $\omega_\odot$ is the apparent angular velocity of the Sun relative to the Earth:
\begin{equation}
    \omega_{\odot} = \frac{2\pi}{365.24 \times 86400} \approx 1.99108 \times 10^{-7} \text{rad s}^{-1},
\end{equation}
and $\lambda$ is the phase shift from the J2000 epoch to the debris generation event. The rotation matrix to convert between the MCI frame and the MCE frame at a given time $t$ is specified by two parameters: the longitude of the ascending node of the lunar equator on the J2000 Ecliptic plane, $\Omega_\textrm{M}(t)$, and the inclination of the lunar equator relative to the J2000 Ecliptic plane, $\varepsilon_\textrm{M}(t)$, which collectively govern the regression of the nodes of the lunar orbital plane (i.e., the intersection with the Ecliptic plane) over a cycle of 18.61 years \cite{rosengren2013long}. Both parameters are obtained from the NASA Jet Propulsion Laboratory's planetary and lunar ephemerides \cite{park2021jpl, williams2008de421}, and the rotation matrix for a given time $t$ is represented as $C_{\textrm{MCE, MCI}}(t) = C_1(\varepsilon_\textrm{M}(t))C_3(\Omega_\textrm{M}(t))$ with $C_1 , C_3$ being the rotation matrices generated by the elementary Euler angle rotations. The Milankovitch orbital element vectors, $\mathbf{h}$ and $\mathbf{e}$, are then defined as: 
\begin{equation}\label{eq: milan_definition}
        \mathbf{h} = \frac{\mathbf{r} \times \mathbf{v}}{\sqrt{\mu a}},\quad 
        \mathbf{e} = \frac{1}{\mu}(\mathbf{v} \times \left(\mathbf{r} \times \mathbf{v})\right) - \frac{\mathbf{r}}{|\mathbf{r}|},
\end{equation}
where $\mu$ is the two-body gravitational parameter and $a$ is the semi-major axis of the RSO. Note that we can normalize the angular momentum vector by $\sqrt{\mu a}$ as for the secular component of any perturbation potential function, the secular change in the semi-major axis is always zero \cite{allan1964long, rosengren2013long}. A given averaged perturbation potential $\overline{\mathcal{R}}$ can then be defined as: 
\begin{equation}
    \overline{\mathcal{R}}(\overline{\mathbf{h}}, \overline{\mathbf{e}}) = \frac{1}{2\pi \sqrt{\mu a}} \int_0^{2\pi} \mathcal{R}(\mathbf{\alpha}, M) \, dM,
\end{equation}
where  $\mathbf{\alpha}$ is an arbitrary set of orbital elements excluding the mean anomaly, $M$ (i.e., the fast-moving variable), and $\overline{\mathbf{h}}$ and $\overline{\mathbf{e}}$ are the average-valued angular momentum and eccentricity vectors with $\overline{\mathcal{R}}(\overline{\mathbf{h}}, \overline{\mathbf{e}})$ independent of the fast-moving variable \cite{rosengren2013long}. The first-order averaged time-evolution of these vectors under $\overline{\mathcal{R}}$ in Lagrangian form is:
\begin{equation}\label{eq: milankovitch_ode}
    \dot{\overline{\mathbf{h}}} = \overline{\mathbf{h}} \times \frac{\partial \overline{\mathcal{R}}}{\partial \overline{\mathbf{h}}} + \overline{\mathbf{e}} \times \frac{\partial \overline{\mathcal{R}}}{\partial \overline{\mathbf{e}}}, \quad 
    \dot{\overline{\mathbf{e}}} = \overline{\mathbf{e}} \times \frac{\partial \overline{\mathcal{R}}}{\partial \overline{\mathbf{h}}} + \overline{\mathbf{h}} \times \frac{\partial \overline{\mathcal{R}}}{\partial \overline{\mathbf{e}}}
\end{equation} 
The solutions to Eq. \eqref{eq: milankovitch_ode} lie along a four-dimensional manifold \cite{tremaine2009satellite}, with two additional constraints on $\overline{\mathbf{h}}$ and $\overline{\mathbf{e}}$: 
\begin{equation}\label{eq: milan_constraints}
    \overline{\mathbf{h}} \cdot \overline{\mathbf{e}} = 0 , \quad \overline{\mathbf{h}}\cdot\overline{\mathbf{h}} + \overline{\mathbf{e}}\cdot\overline{\mathbf{e}} = 1,
\end{equation}
thus reducing the system from six to four degrees of freedom. Since all analyses in this work utilizing the Milankovitch orbital element vectors focus solely on the average-valued vectors, we henceforth drop the overbar notation for simplicity, and denote $\bar{\mathbf{h}}$ simply as $\mathbf{h}$ and likewise for $\bar{\mathbf{e}}$.

\section{Probabilistic Modeling of Short Time Horizon Outcomes}\label{sec: modeling_sinking}

To assess the probability of various debris outcomes in a more rigorous manner, it is important to consider the probability space of debris fragments that could be generated at a given originating location by the SBM, and subsequently to characterize the variations in the probability of each outcome across a wide gamut of such locations. The approach proposed in this section provides a measure-theoretic basis for the consideration of the probability of fragments sinking to the lunar surface over a short time horizon (i.e., a fragment's orbital period, typically on the order of hours). For each fragment, we define this time horizon as their osculating two-body orbital period at the time of the debris generation event: 
\begin{equation}
    T = 2\pi \sqrt{\frac{a^3}{\mu}}
\end{equation}
This time horizon is chosen to characterize the probability of sinking since we may leverage the fact that if a fragment has a perilune distance $r_\textrm{p}$ below the lunar surface \textit{at the time of generation} then it will sink within the first fragment period after generation; enabling a prediction of whether a given fragment sinks without having to propagate its trajectory. The theoretical development of the probabilistic framework for the short time horizon outcomes is presented in Section \ref{sec: sink_theory}, deriving the required distributions for one originating location as well as for the variations across locations. The fitting of relevant distributions is then briefly discussed in Section \ref{sec: sinking_fitting}, followed by the definition of the Monte Carlo baseline in \ref{sec: mc_baseline} demonstrating that the static perilune threshold matches well with the outcomes from the propagation of trajectories. Finally, Section \ref{sec: comparison_sinking} provides a comparison between the proposed framework and the Monte Carlo baseline, illustrating the accuracy of resulting probability of sinking across originating locations in the LLO domain. 

\subsection{Theoretical Development of Short Time Horizon Framework}\label{sec: sink_theory}

Consider the space of all possible $\mathbf{\Delta v}$ values that could be generated by the SBM, represented as a probability space $(\Omega_{\mathbf{\Delta v}}, \mathcal{F}_{\mathbf{\Delta v}}, \mathbb{P}_{\mathbf{\Delta v}})$ with the sample space $\Omega_{\mathbf{\Delta v}} = \mathbb{R}^3$, the associated event space $\mathcal{F}_{\mathbf{\Delta v}} = \mathcal{B}(\mathbb{R}^3)$ i.e., the Borel sigma algebra over $\mathbb{R}^3$, and the probability measure $\mathbb{P}_{\mathbf{\Delta v}}$. The samples generated by this space follow the SBM's sampling approach for the magnitude and direction as defined in Eq. \eqref{eq:marginals_dv}. For a  given fragment, the semi-major axis $a$ as a function of its $\mathbf{\Delta v}$ is parameterized by the originating location $(\mathbf{r}_o, \mathbf{v}_o)$ by equating the $|\mathbf{r}_o|$ term of the vis-viva equation for the originating RSO and the fragment: 
\begin{equation}
    a(\mathbf{\Delta v} ; \mathbf{r}_o, \mathbf{v}_o) = \frac{\mu a_o}{\mu - a_o (2 \mathbf{v}_o \cdot \mathbf{\Delta v} + \mathbf{\Delta v \cdot \Delta v})} ,
\end{equation}
where $a_o$ is the semi-major axis of the originating RSO's orbit, as recovered from $(\mathbf{r}_o, \mathbf{v}_o)$. The standard definition of the eccentricity vector with $\mathbf{v} = \mathbf{v}_o + \mathbf{\Delta v}$, gives the fragment's eccentricity $e$ as:
\begin{equation}
    \begin{split}
        \mathbf{e}(\mathbf{\Delta v} ; \mathbf{r}_o, \mathbf{v}_o) &= \frac{1}{\mu} \left( \left[(\mathbf{v}_o + \mathbf{\Delta v})\cdot(\mathbf{v}_o + \mathbf{\Delta v})  - \frac{\mu}{|\mathbf{r}_o|} \right]\mathbf{r}_o - [\mathbf{r}_o \cdot (\mathbf{v}_o + \mathbf{\Delta v})] (\mathbf{v}_o + \mathbf{\Delta v})\right) \\
        e(\mathbf{\Delta v} ; \mathbf{r}_o, \mathbf{v}_o) &= |\mathbf{e}(\mathbf{\Delta v} ; \mathbf{r}_o, \mathbf{v}_o)|
    \end{split}
\end{equation}
These functions enable a definition of the perilune distance attained by a fragment with a given $\mathbf{\Delta v}$ as:
\begin{equation}
    r_{\textrm{p}}(\mathbf{\Delta v} ; \mathbf{r}_o, \mathbf{v}_o) = a(\mathbf{\Delta v} ; \mathbf{r}_o, \mathbf{v}_o)(1 - e(\mathbf{\Delta v} ; \mathbf{r}_o, \mathbf{v}_o))
\end{equation}
Then, for a measurable space over all possible perilune distances, denoted as $(E_{r_\textrm{p}}, \mathcal{E}_{r_\textrm{p}}) = (\mathbb{R}, \mathcal{B}(\mathbb{R}))$, this function may consequently be thought of as a mapping from $(\mathbb{R}^3, \mathcal{B}(\mathbb{R}^3))$ to $(\mathbb{R}, \mathcal{B}(\mathbb{R}))$. For any  measurable Borel set $B \subset  \mathcal{E}_{r_\textrm{p}}$, the probability distribution over the perilune distances is the push-forward measure of $\mathbb{P}_{\mathbf{\Delta v}}$ under the mapping $r_\textrm{p}$:
\begin{equation}
    (r_{\textrm{p}\,\star}\mathbb{P}_{\mathbf{\Delta v}})(B) = \mathbb{P}_{\mathbf{\Delta v}}(r_\textrm{p}^{-1} (B)),
\end{equation}
where the subscript $\star$ denotes the push-forward measure.

The probability distribution $\mathbb{P}_{\mathbf{\Delta v}}$ can thus be used to obtain a probability distribution over the perilune distances for a given originating location, finding the binary outcome of whether a fragment sinks (i.e., $r_\textrm{p} \leq r_\textrm{M}$, the lunar radius) by defining an indicator function to map the continuous range of perilunes to $\{ 0 , 1\} $ with $0$ representing non-sinking fragments and $1$ representing sinking fragments. We first define a `sinking set':
\begin{equation}
   \Omega_{\textrm{sink}} = \left\{ \mathbf{\Delta v} \in \mathbb{R}^3  \mid  r_\textrm{p}(\mathbf{\Delta v} ; \mathbf{r}_o, \mathbf{v}_o) \leq r_\textrm{M} \right\}, 
\end{equation}
and then write the indicator function as $\mathbf{1}_{\Omega_{\textrm{Sink}}}$ which can be used as a push-forward measure on $\mathbb{P}_{\mathbf{\Delta v}}$ to obtain a probability distribution over a Bernoulli random variable $ S = \mathbf{1}_{\Omega_{\textrm{Sink}}} (\mathbf{\Delta v}) : \mathbb{R}^3 \rightarrow \{ 0, 1 \}$ with $ \mathbb{P}_{S} = \mathbf{1}_{\Omega_\textrm{sink} \, \star}\mathbb{P}_{\mathbf{\Delta v}}$.
Then, the probability of sinking i.e., $\mathbb{P}_{S}(S = 1)$ is the evaluation of this push-forward measure on the set $\{ 1\}$: 
\begin{equation}
\label{eq:bern_integral}
   \mathbb{P}_S(S = 1) = \mathbb{P}_{\mathbf{\Delta v}}(\mathbf{1}_{\Omega_\textrm{sink}}^{-1}(\{1\})) = \int_{\Omega_{\textrm{sink}}} \rho_{\mathbf{\Delta v}}(\mathbf{\Delta v}) d\mathbf{\Delta v} \, ,
\end{equation}
where $\rho_{\mathbf{\Delta v}}$ is the probability density of $\mathbb{P}_{\mathbf{\Delta v}}$.
The probability obtained in Eq. \eqref{eq:bern_integral} corresponds to a single Bernoulli trial (i.e., one fragment). For a debris generation event with $N$ fragments and each fragment $i$ with $\mathbf{\Delta v}_i$, consider the associated random variables $S_i  = \mathbf{1}_{\textrm{sink}}(\mathbf{\Delta v}_i) \in \{0, 1\}$, where each $S_i$ is an independent and identically distributed (i.i.d) Bernoulli trial. Then, the total number of sinking fragments can be calculated as $S_N = \Sigma_{i=1}^N S_i$. Denoting the resulting probability from the evaluation of the integral in Eq. \eqref{eq:bern_integral} as $p_s$, we equivalently write $S_N$ as: 
\begin{equation}\label{eq: binom}
\begin{split}
    S_N &\sim \text{Binomial}(N, p_s), \quad \mathbb{P}(S_N = n) = \binom{N}{n} p_s^n (1-p_s)^{N-n}
\end{split}
\end{equation}
Lastly, to more directly compare the results of this approach with the empirical fractions employed by existing literature \cite{boone2023simulation}, we convert the number of sinking fragments represented by Eq. \eqref{eq: binom} to a fraction $\gamma_s \in [0, 1]$: 
\begin{equation}\label{eq: gamma}
    \Gamma_s = \frac{S_N}{N}, \quad 
    \mathbb{P}(\Gamma_s = \gamma_s) = \mathbb{P}(S_N = N\gamma_s) = \binom{N}{N\gamma_s}p_s^{N\gamma_s} (1-p_s)^{N - N\gamma_s}
\end{equation}
Note that the expected value for the fraction of sinking fragments under this formulation is exactly the probability of sinking for one Bernoulli trial, demonstrating that the Binomial probability distribution is an unbiased estimator of the probability of sinking across a set of fragments of any desired size. This formulation can further be extended across originating locations to assess how the probability defined in Eq. \eqref{eq:bern_integral} varies as we change $(\mathbf{r}_o, \mathbf{v}_o)$. The development of such a `map' of the sinking probability operationalizes the short time horizon component of probabilistic framework, providing mission designers the ability to rapidly survey regions of interest within the LLO regime for debris mitigation considerations during preliminary orbit selection. Consider the Keplerian representation of an originating location: 
\begin{equation}
\renewcommand{\arraystretch}{0.30}
      \mathbf{oe}  = \begin{bmatrix}
        a, & e, & i, & \Omega, & \omega, & \nu
    \end{bmatrix}^T = g(\mathbf{r}, \mathbf{v}), \quad (\mathbf{r}, \mathbf{v}) = g^{-1}(\mathbf{oe}),
\end{equation}
where $e$ is the eccentricity, $i$ is the inclination, $\Omega$ is the right ascension of the ascending node, $\omega$ is the argument of perilune, and $\nu$ is the true anomaly. For any state $(\mathbf{r}, \mathbf{v})$ represented in Cartesian coordinates, $\mathbf{oe} = g(\mathbf{r}, \mathbf{v})$ can be recovered using the standard relationships between Cartesian and Keplerian elements \cite{prussing1993orbital}, and vice versa using $g^{-1}$. Then, for a given orbital plane (i.e., holding $i, \Omega, \omega$ constant) and true anomaly, each originating location can be equivalently denoted as a pair $(a_i, e_j)$ which can be used to recover the Cartesian representation exactly, and the sinking set for the $(i,j)^{th}$ originating location is written as as:
\begin{equation}
    \Omega_{\textrm{sink}, i, j} = \left\{ \mathbf{\Delta v} \in \mathbb{R}^3  \mid (\mathbf{r}_{ij}, \mathbf{v}_{ij}) = g^{-1}(a_i, e_j ; i, \Omega, \omega, \nu) \, ,  r_\textrm{p}(\mathbf{\Delta v} ; \mathbf{r}_{ij}, \mathbf{v}_{ij}) \leq r_\textrm{M} \right\}
\end{equation}
The probability of sinking for a given originating location is therefore: 
\begin{equation}\label{eq: sink_per_loc}
    p_s(a_i, e_j) = \int_{\Omega_{\textrm{sink}, i, j}} \rho_{\mathbf{\Delta v}}(\mathbf{\Delta v}) d\mathbf{\Delta v} \, , 
\end{equation}
which can be substituted into Eq. \eqref{eq: gamma} as necessary. Note that while we have formulated the set of originating locations in terms of the semi-major axis and eccentricity for intuitive utilization during preliminary orbit selection, any subset of the Keplerian orbital elements (or the entire Cartesian state) may be used to investigate more complex design spaces. 

\subsection{Fitting of Relevant Distributions for Short Time Horizon Framework}\label{sec: sinking_fitting}

\begin{figure}[h!]
    \centering
    \includegraphics[width=0.55\linewidth]{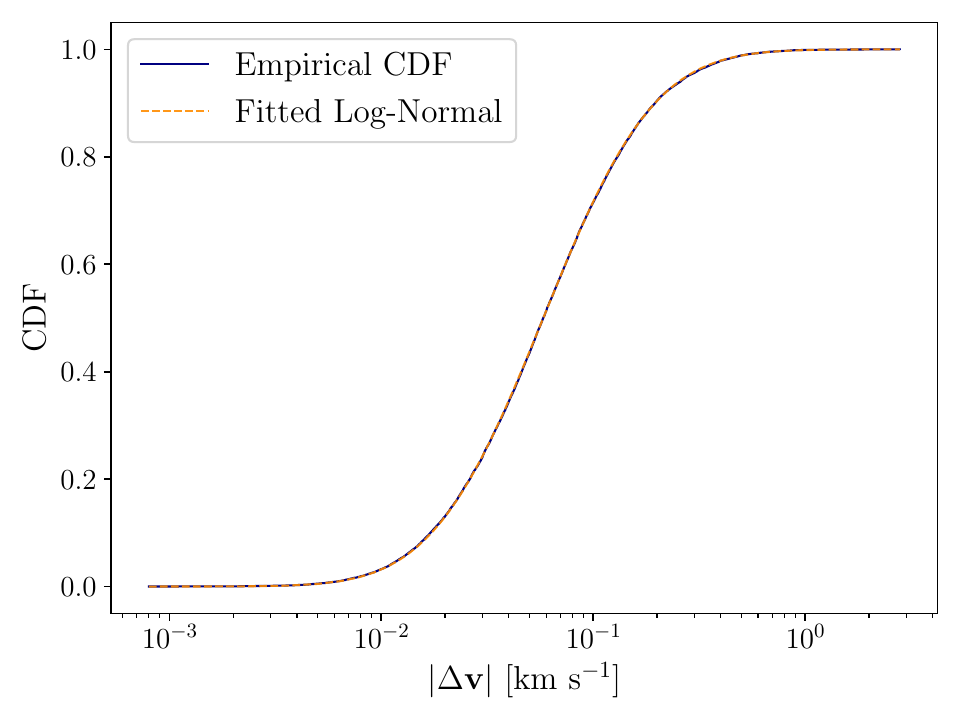}
    \caption{Plot of Empirical CDFs and the CDF of the fitted distribution for $\rho_{|\mathbf{\Delta v}|}$, using a sample with $n = 9500$, with a resulting $\mathbb{P}(D_n > d_n) = 0.968$ and $d_n = 0.004$.}
    \label{fig:fit_dv}
\end{figure}

To implement the short time horizon component of the proposed framework, we must know the probability density $\rho_{\mathbf{\Delta v}}$. Since $|\mathbf{\Delta v}|$ is sampled independently from the uniformly-sampled direction components $(u, \theta)$, as defined in  Eq. \eqref{eq:marginals_dv}, a univariate probability distribution is fitted to the $|\mathbf{\Delta v}|$ values for a generated sample of $2 \times 10^{6}$ fragments from the NASA SBM, setting the model parameters such that $L_{c} \in [0.01 \text{ m}, 1\text{ m}]$. The fitting of all relevant distributions for this work is conducted using the OpenTURNS software package \cite{baudin2017open} in Python, and the same is additionally used to assess the goodness of fit for the resulting distributions. To find the best-fit distribution in each case, we consider various named distributions (e.g., Gaussian, Beta, etc.), denoted as $\mathcal{M}_i$, each with $\beta_i$ associated parameters. Then, for a sample of size $n$, we denote the maximum likelihood of that sample with respect to $\mathcal{M}_i$ as $\mathcal{L}_i$. The best-fit distribution is then the one which minimizes the Bayesian Information Criterion $\text{BIC}(\mathcal{M}_i)$: 
\begin{equation}\label{eq: BIC}
        \mathcal{M}_{\text{BIC}} = \arg \min_{\mathcal{M}_i} \text{BIC}(\mathcal{M}_i), \quad \text{BIC}(\mathcal{M}_i) = -2 \frac{\log \mathcal{L}_i}{n} + \frac{\beta_i \log n}{n},
\end{equation}
where the first term maximizes the log likelihood and the second term penalizes overfitting through a large number of parameters. The $\mathcal{M}_{\text{BIC}}$ for $\rho_{|\mathbf{\Delta v}|}$ is found to be a log-normal distribution with parameters $\mu_{\log, |\mathbf{\Delta v}|} = -2.84434$ and $\sigma_{\log, |\mathbf{\Delta v}|} = 0.946695$. Both parameters converge to these values as the sample size $n$ is increased, and stay largely constant for $n \geq 2 \times 10^5$. The Kolmogorov-Smirnov one-sample test is  used to confirm the goodness of fit of this distribution, assessing whether a sample of data is drawn from a given probability distribution. If we denote the (unknown) cumulative density function (CDF) from which the empirical sample is drawn as $F$, and the CDF of the fitted distribution as $G$, then the null hypothesis becomes $\mathcal{H}_0 = \{ F = G\}$. Defining $X_1 \dots X_n$ as independent random variables drawn from $F$, the empirical CDF is: 
\begin{equation}
    F_n(x) = \frac{1}{n}\sum_{i=1}^n \mathbf{1}_{X_i \leq x} \,,  \forall x \in \mathbb{R}
\end{equation}
The Kolmogorov-Smirnov (K-S) test statistic is then defined as $D_n = \sqrt{n} \sup_x |F_n(x) - G(x)|$, and the empirical value of the test statistic, $d_n$ using the realization of $F_n$ on the sample. Fixing a risk threshold $\delta$ (typically 0.05), we accept the null hypothesis if $\mathbb{P}(D_n > d_n) \geq \delta$. As can be seen in Fig. \ref{fig:fit_dv}, the empirical and fitted CDFs show good visual agreement for a sample of size $n = 9500$ (typical for one debris generation event with the given $L_{c, \min} = 0.01$ m) and the $p-$value i.e., $\mathbb{P}(D_n > d_n) = 0.968 \gg \delta = 0.05$, thus confirming the goodness of fit of the log-normal distribution through the acceptance of the null hypothesis of the Kolmogorov-Smirnov test. 

\subsection{Monte Carlo Baseline}\label{sec: mc_baseline}

To validate the results of the probabilistic framework for the short time horizon assessment of sinking fragments, a Monte Carlo baseline that is similar to the approaches used by existing literature must be defined. For a debris generation event at a given originating location with $N$ generated fragments, the fraction of sinking fragments $f_s$ and non-sinking fragments $f_{ns}$ are defined as: 
\begin{equation}\label{eq: sinking_empirical}
    f_s = \frac{\text{Number of fragments with } r_\textrm{p} \leq r_\textrm{M}}{N}, \quad f_{ns} = 1 - f_s
\end{equation}
with $f_s$ serving as the empirical analogue to the analytical $p_s$. To ensure that these fractions accurately capture debris behavior, we must propagate generated debris fragments and verify that the propagated outcomes under the high-fidelity dynamics described in Eq. \eqref{eq: dynamics_overall} match those predicted by these fractions. For this purpose, we use a circular polar LLO at an altitude of 100 km with the following Keplerian orbital elements in the MCI frame: 
\begin{equation}
    a = 1838.1 \text{ km}, e= 0, i = 90^\circ , \Omega = 0^\circ, \omega = 270^\circ,
\end{equation}
and identify ten originating locations along the orbit which are evenly-spaced in time. At each originating location, debris generation events are simulated at uniformly sampled epochs between 11:59:00 1 January 2025 to 11:59:00 1 January 2026 to obtain $\approx10^5$ fragments at each originating location. The parameters of the SBM are set such that $L_c \in [0.01 \text{m}, 1.0 \text{m}]$, and the fragment trajectories are propagated using NASA's GMAT \cite{hughes2016general} via the Python API \cite{conway2020using} with all forces as described in Section \ref{sec: dynamics}. Table \ref{tab:miscateg} below shows the mean mis-categorization between the predicted and propagated outcomes across the Monte Carlo runs for all originating locations. 

\begin{table}[h!]
    \centering
    \caption{Percentage of predicted (rows) outcomes resulting in propagated (columns) outcomes}
    \begin{tabular}{|c|c|c|}
        \hline
         & \textbf{Sinking} & \textbf{Non-Sinking} \\
         \hline
         \textbf{Sinking} & $99.546$ & $0.454$\\
         \hline
         \textbf{Non-Sinking} & $0.106$ & $99.894$\\
         \hline
    \end{tabular}
    \label{tab:miscateg}
\end{table}

\begin{figure}[h!]
    \centering
    \includegraphics[width=\linewidth]{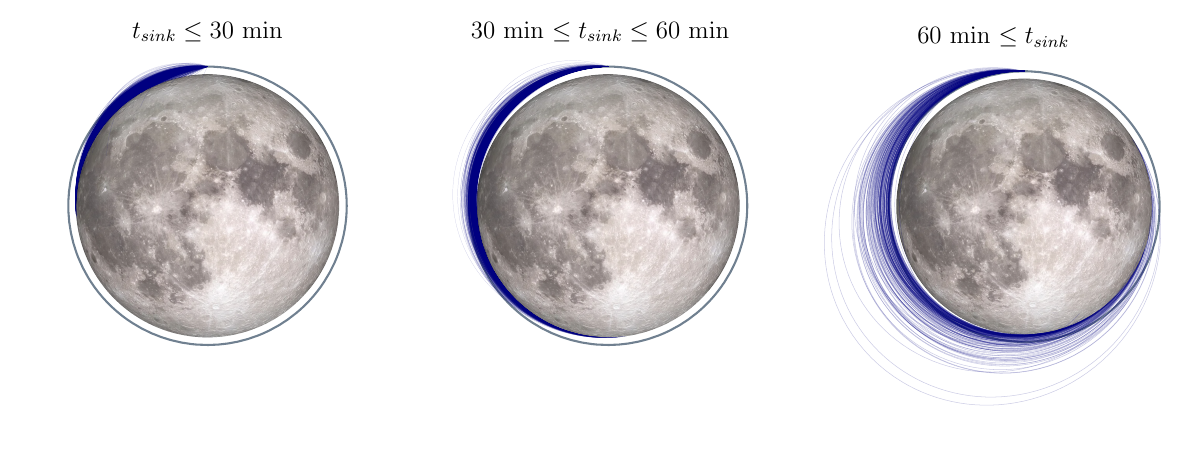}
    \caption{Trajectories (navy) of predicted sinking fragments from a debris generation event in a 100 km circular polar LLO (gray), visualized in the $\hat{i}$-$\hat{k}$ MCI plane across three ranges of sinking time $t_{sink}$.}
    \label{fig:sinking_traj}
\end{figure}

As can be seen, the static prediction of sinking and non-sinking outcomes using the fragment perilune distances at the time of the debris generation event is an accurate representation of the realized outcome upon propagation of trajectories. The small percentage of mis-categorizations may be attributed to the effects of the perturbative forces included in the propagation, which are not captured in the two-body assumption used by these empirical fractions. We can further visually inspect the trajectories of the sinking fragments to qualitatively assess their behavior, as shown in Fig. \ref{fig:sinking_traj}. The originating orbit (100 km circular polar LLO) is  shown in gray, and the trajectories are separated based on the time $t_{sink}$ from the generation of the fragment to the lunar impact. 

Some fragments can be seen to sink almost immediately after the debris generation event, while others (depending on the specific $\mathbf{\Delta v}$ value) may take almost the entire fragment period to sink to the lunar surface. Note that while all of these fragments sink within a few hours after the debris generation event, they may pass through orbital altitudes on the order of thousands of kilometers above the lunar surface during the first fragment period before ultimately sinking; this further motivates future work towards the development of environmental flux models analogous to ORDEM in the near-Earth case, such that risks during the immediate time after a debris generation event may be rigorously modeled. 

\subsection{Assessment of Sinking Fragments}\label{sec: comparison_sinking}

\begin{figure}[h!]
    \centering
    \includegraphics[width=\linewidth]{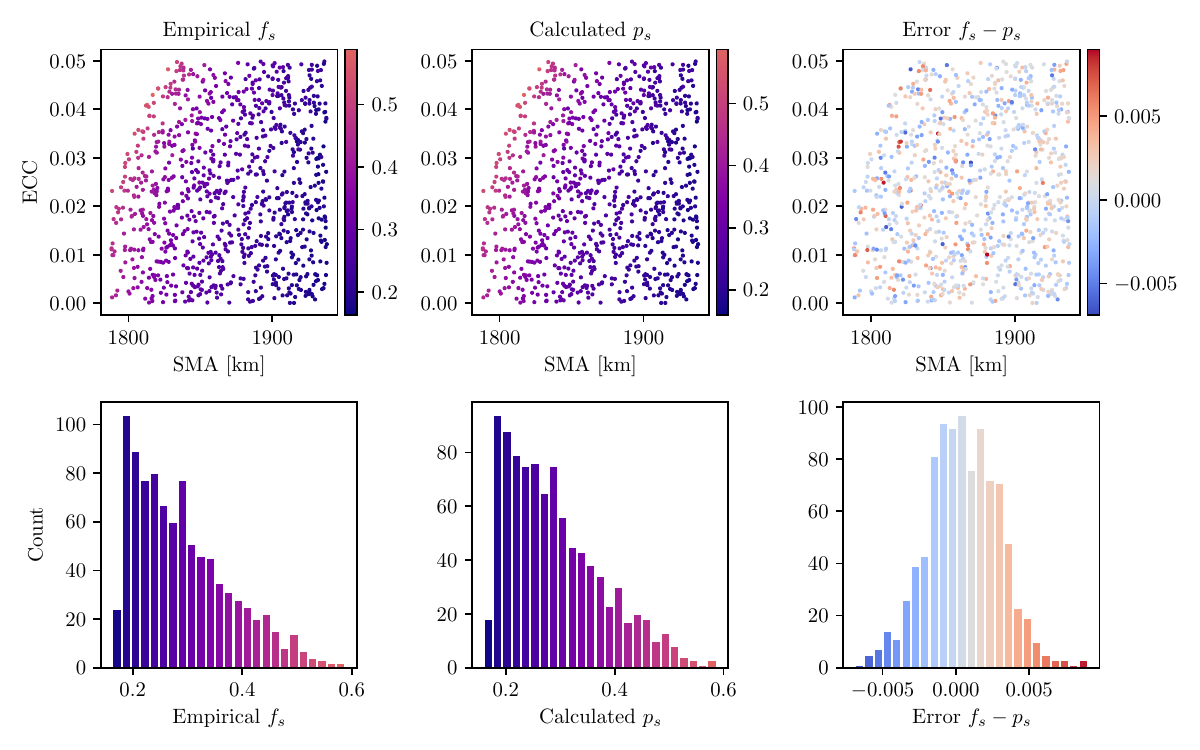}
    \caption{Comparison of analytical sinking probability $p_s$ with empirical sinking fraction $f_s$.}
    \label{fig:sinking_main_result}
\end{figure}

We can now use the fitted distributions to compare the results of the probabilistic framework for short time horizon debris outcomes with those from the Monte Carlo baseline. To do so, we consider the orbital plane with $i = 90^{\circ}, \Omega = 0^{\circ}, \omega = 270^{\circ}$ (i.e., containing the polar LLOs) and uniformly sample 5,000 pairs of semi-major axis and eccentricity $(a_i, e_j)$ with $a_i$ between 50 km and 250 km above the lunar surface, and $e_j \in [0.0, 0.05]$. Then, for each pair, five true anomaly values are additionally sampled uniformly to define five originating locations on each orbit specified by the pair. A debris generation event is simulated at each location, with the SBM parameters set to sample $L_c \in [0.01 \text{ m}, 1 \text{ m}]$, and the empirical sinking fraction $f_s$ is calculated as described in Eq. \eqref{eq: sinking_empirical}. The average value of $f_s$ across all five locations on the orbit is used as the representative value of the empirical sinking probability for the $(a_i, e_j)$ pair. Likewise, the analytical probability $p_s$ is calculated for each originating location using Eq. \eqref{eq: sink_per_loc} as implemented using the OpenTURNS package \cite{baudin2017open}, with the results for averaged values of $f_s$ and $p_s$ shown in Fig. \ref{fig:sinking_main_result}.

The resulting probabilities from the proposed framework show close alignment with the empirical sinking fractions across all originating locations considered. Further, the errors can be seen to be on the order of $10^{-2}$ and are uncorrelated with the underlying $(a_i, e_j)$ pair. The likelihood of sinking seen to be highest for originating locations where the perilune distance of the originating location itself is close to the lunar radius, where smaller $|\mathbf{\Delta v}|$ values are sufficient to cause the generated fragments to sink. Thus, originating locations with smaller semi-major axis values and higher eccentricity values may see a majority of generated fragments sinking within one fragment period, which may be desirable to mission designers aiming to mitigate debris risks to proximate missions. Further discussion of the utility of this framework for to mission designers for preliminary orbit selection, as well as to policymakers for investigating useful thresholds for the minimum sinking probability are briefly presented in Section \ref{sec: results_utilization}. 

\section{Probabilistic Modeling of Medium Time Horizon Outcomes}\label{sec: modeling_nonsinking}

As defined in the previous section, the set of debris fragments generated for a given on-orbit breakup event in the LLO regime will sink to the lunar surface within the first fragment period with probability $p_s$. The remaining fragments will linger in the LLO regime over longer time horizons, with probability $1-p_s$, and will have trajectories that are influenced by the various perturbative forces described in Section \ref{sec: dynamics}. To assess the effects of these forces over longer time-horizons on the outcomes of the `non-sinking' fragments, we are primarily interested in identifying both the percentage of any additional sinking fragments from this non-sinking set as well as the physical characteristics of these fragments. To model the behavior of these non-sinking fragments over time-horizons spanning days to months, the relative magnitude of the various perturbative forces that may affect their trajectories is first characterized in Section \ref{sec: dyn_rel_mag} and the solar radiation pressure is identified as a dominant perturbation of interest. Next, the time-evolution of the average-valued Milankovitch orbital element vectors introduced in Section \ref{sec: milankovitch_background} under the averaged perturbation potential due to SRP is derived in Section \ref{sec: milankovitch_evolution}. Further, given the dependence of SRP on the fragment's area-to-mass ratio $\psi$, a conditional distribution $\rho_{\psi | S = 0}$ over the area-to-mass ratio values for the non-sinking fragments is formulated in Section \ref{sec: nonsink_cond_amr}, which is subsequently used to define the initial distribution of the Milankovitch orbital element vectors. The implementation of fitted distributions for the probabilistic formulation is discussed in Section \ref{sec: fitting_nonsink}, with comparisons of the results to a Monte Carlo baseline presented in Section \ref{sec: comparison_nonsinking}. 

\subsection{Relative Magnitude of Perturbative Forces}\label{sec: dyn_rel_mag}

To consider the effects of various perturbative forces on the the behavior of non-sinking fragments, the relative magnitude of each source of perturbation is first characterized. While the aspherical lunar gravity and third-body effects from the Sun and the Earth are only dependent on the fragment's instantaneous position $\mathbf{r}$, the SRP additionally depends on the fragment's area-to-mass ratio $\psi$, and will thus have a variable range of magnitudes across the generated fragments with the same position. To facilitate the comparison of relative magnitudes, the SRP with $\psi = 1$ is evaluated; while $\psi = 1$ is greater than the mean area-to-mass ratio value for the non-sinking fragments, as discussed further in Section \ref{sec: fitting_nonsink}, the dependence of acceleration due to SRP on $\psi$ is linear, and the `reference' case of $\psi = 1$ can be scaled by any given $\psi$ value. 

\begin{figure}[h!]
    \centering
    \begin{subfigure}[b]{0.45\textwidth}
         \centering
         \includegraphics[width=\textwidth]{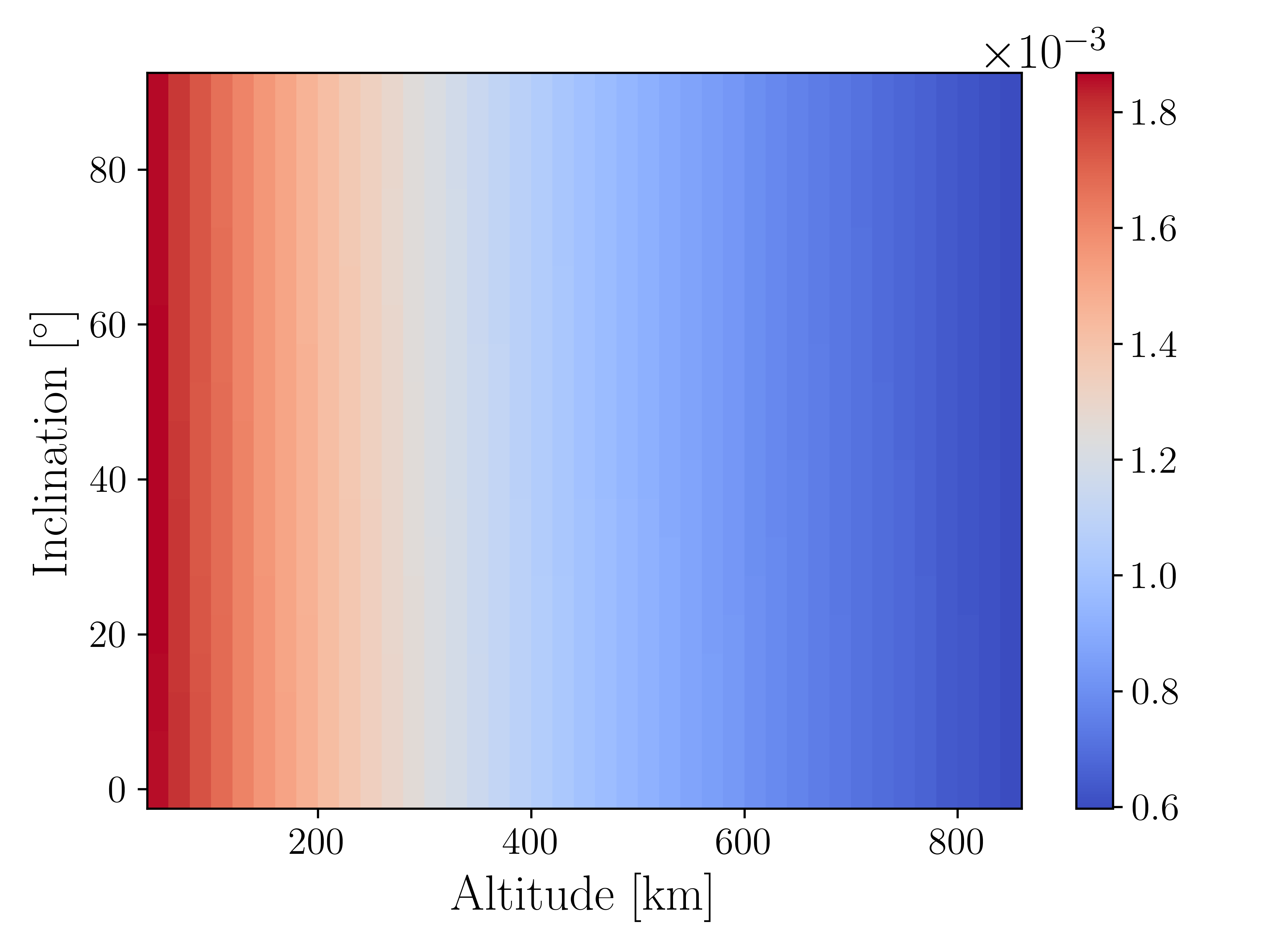}
         \caption{Asph. Gravity ($L = 100$)}
         \label{fig:asph_ord_100}
     \end{subfigure}
     \begin{subfigure}[b]{0.45\textwidth}
         \centering
         \includegraphics[width=\textwidth]{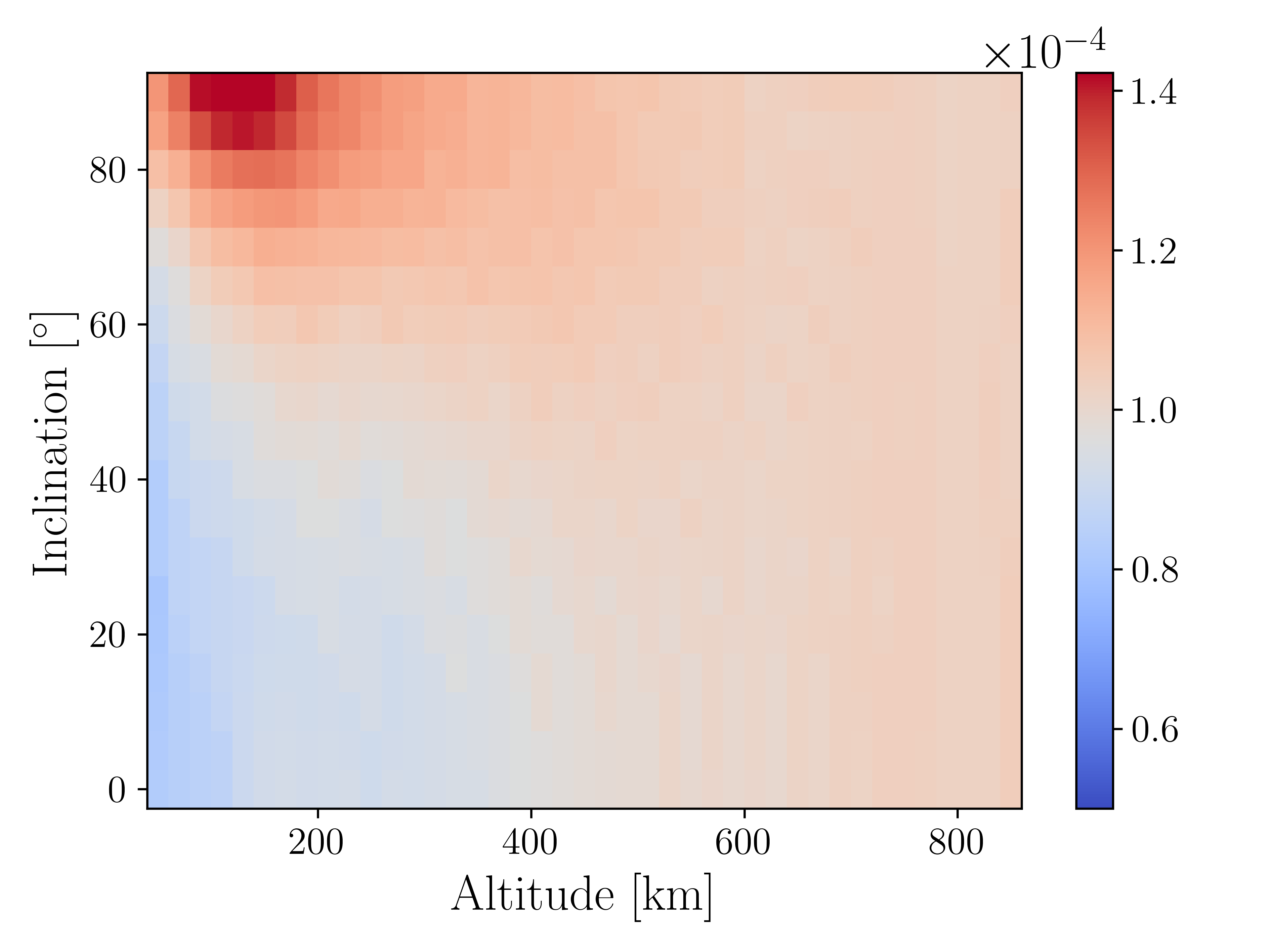}
         \caption{SRP ($\psi = 1$)}
         \label{fig:srp_one_period}
     \end{subfigure}
    \caption{Mean acceleration magnitudes for aspherical lunar gravity and SRP  over one orbital period.}
    \label{fig:rel_mag_one_period}
\end{figure}

A grid over altitudes (ranging from 50 km to 850 km above the lunar surface at intervals of 20 km) and inclinations (ranging from $0^\circ$ to $90^\circ$ at intervals of $5^\circ$) is considered, with a circular LLO at each-grid point serving as the representative initial condition for propagation\footnote{Visual artifacts, especially in any sharp variation of acceleration magnitudes in neighboring grid cells, may be a result of the resolution of the grid. Finer grids may ameliorate the assessment of precise acceleration magnitudes for an LLO sub-region of interest; the primary goal of the grid used here is to gain insights into the relative strength of perturbative sources across a broad region of the LLO environment.}. The initial condition is propagated under the dynamics specified in Eq. \eqref{eq: dynamics_overall} for one orbital period for each location on the grid using the epoch 11:59:00 1 January 2026, and the accelerations due to individual perturbations are obtained at each point on the state history using GMAT \cite{hughes2016general}. The mean perturbative acceleration magnitude $\overline{|\ddot{\mathbf{r}}_{\textrm{Perturbation}}|}$ over the entire orbit is used as a representative value for that perturbative source at the given location. As is expected, the two-body acceleration is the dominant force across the regime considered, with magnitudes ranging from $0.5$ to $1.5$ m s$^{-2}$ across the considered locations. The third-body effects from the Sun and the Earth combined have the lowest acceleration magnitudes of all perturbations, staying on the order of $10^{-5}$ m s$^{-2}$ with the strongest effects at higher altitudes. The primary comparison is thus focused on the perturbations due to aspherical lunar gravity and SRP. Figure \ref{fig:rel_mag_one_period} shows the magnitude of the acceleration due to these sources over the grid of LLO locations. 

As shown in Fig. \ref{fig:asph_ord_100}, the perturbative effects due to aspherical lunar gravity (using $L=100$ in Eq. \eqref{eq: spherical_harmonics}, consistent with those used by Boone et al. \cite{boone2023simulation}), result in mean acceleration magnitudes on the order of up to $10^{-3}$ m s$^{-2}$, with the strongest effects being at altitudes close to the lunar surface. As the altitude increases, the mean acceleration magnitude due to these terms reduces to the order of $10^{-4}$ m s$^{-2}$. In comparison, the effects of SRP (using $\psi = 1$) are on the order of $10^{-4}$ m s$^{-2}$ across the grid of locations considered, as shown in Fig. \ref{fig:srp_one_period}. At the considered epoch, the Earth is near its perihelion, and the mean sunlit fraction ($\overline{1 - \nu}$) ranges between 0.6 to 0.8, depending on how the orbit at each location moves through the shadow of the Moon and the Earth. For the values of $\psi \in [10^{-2}, 10^1]$ m$^2$ kg$^{-1}$ typically generated by the SBM as shown in Fig. \ref{fig:debris_char}, the magnitude of acceleration due to SRP correspondingly ranges between $10^{-6}$ to $10^{-3}$ m s$^{-2}$. Thus, for most fragments with smaller $\psi$ values, the dominant averaged perturbative effect over short time horizons would typically be $\ddot{\mathbf{r}}_{\textrm{Asph. Gravity}}$. 

\begin{figure}[h!]
    \centering
     \begin{subfigure}[b]{0.45\textwidth}
         \centering
         \includegraphics[width=\textwidth]{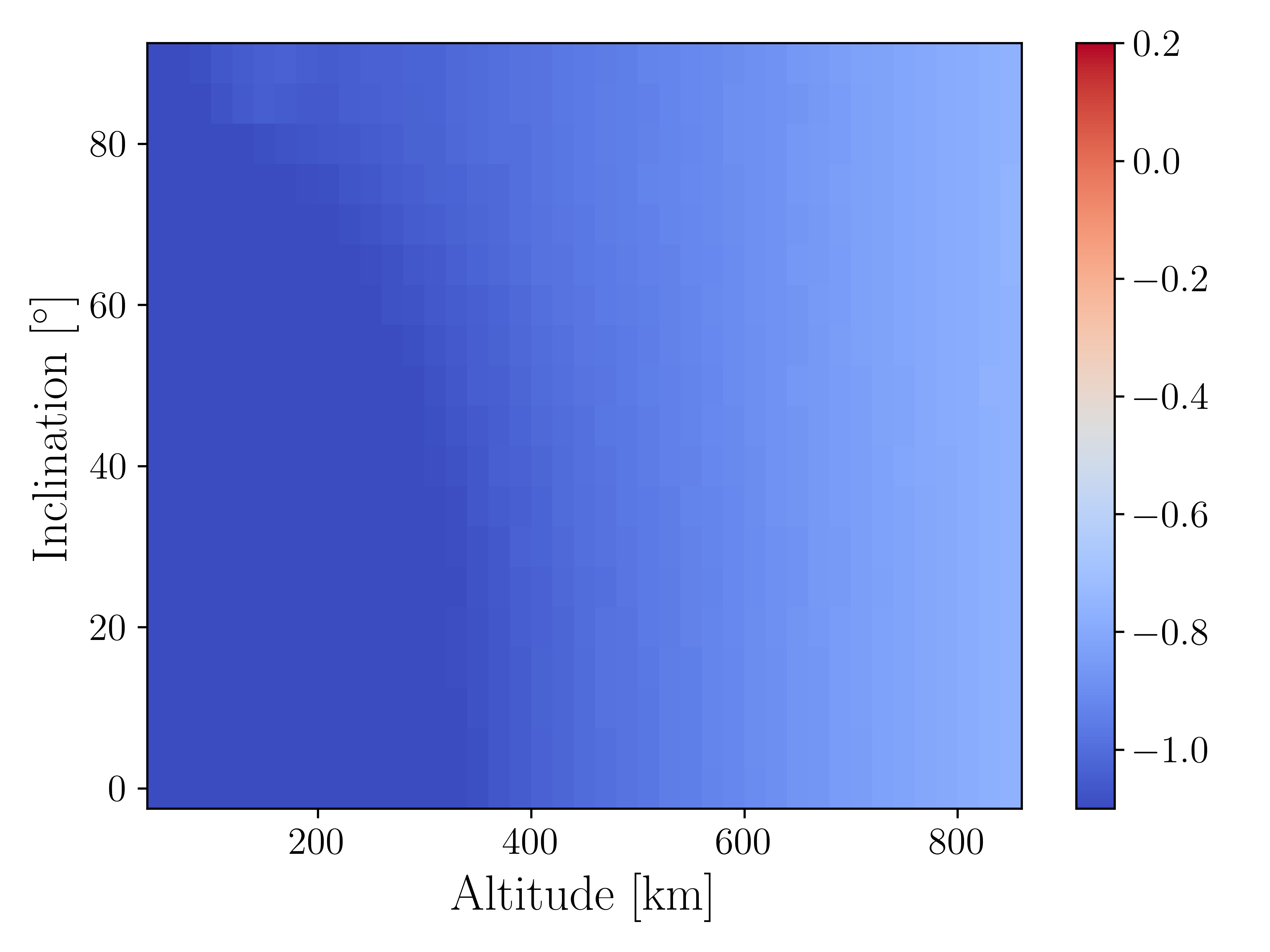}
         \caption{One period case}
         \label{fig:comp_one_period}
     \end{subfigure}
     \hfill    
     \begin{subfigure}[b]{0.45\textwidth}
         \centering
         \includegraphics[width=\textwidth]{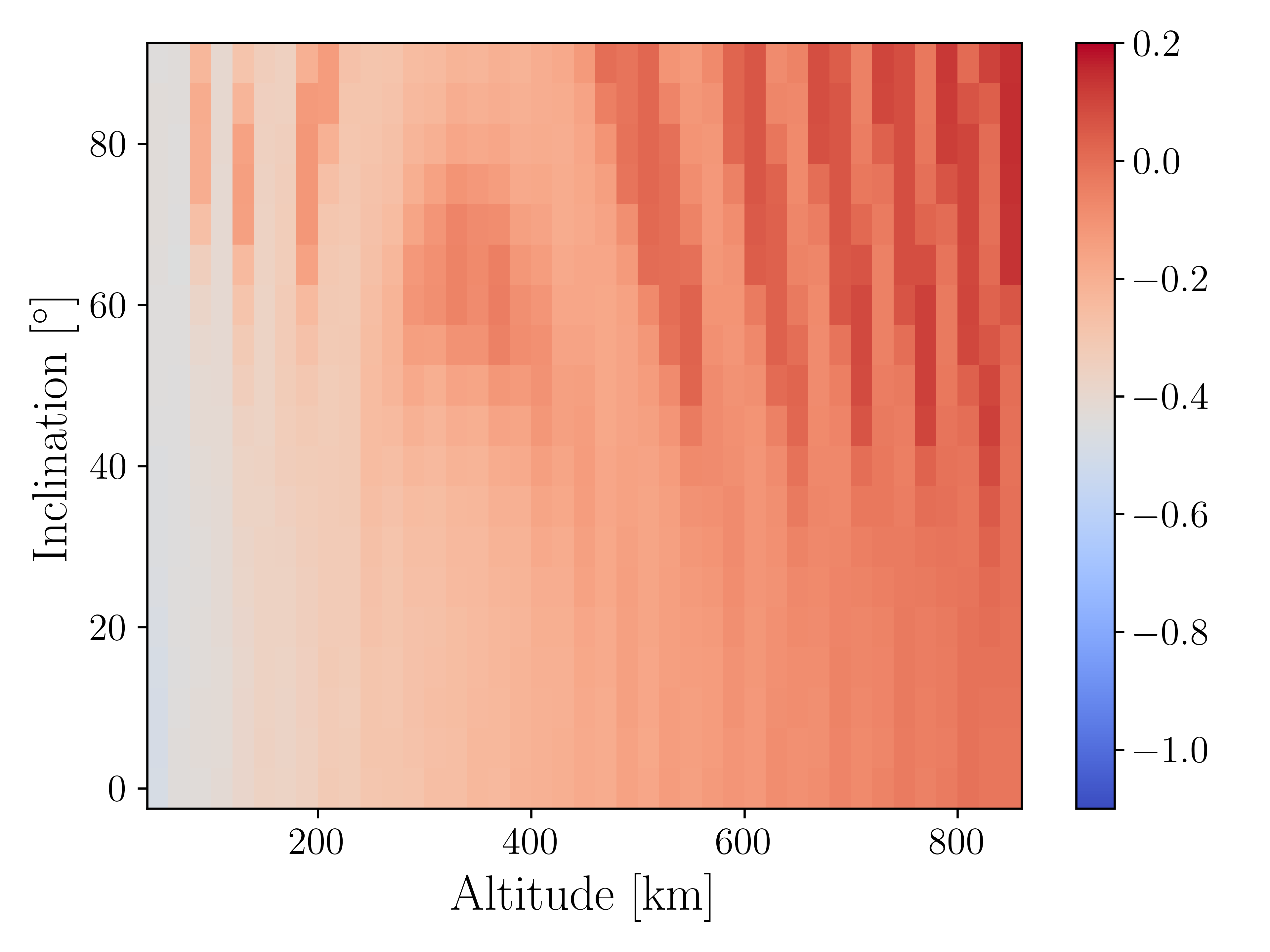}
         \caption{One hundred period case}
         \label{fig:comp_hundred_periods}
     \end{subfigure}
    \caption{Log-relative-magnitude comparison of SRP and Asph. Gravity over one period (left) and one hundred periods (right).}
    \label{fig:log_rel_mag}
\end{figure}

However, the mean acceleration magnitudes due to SRP tend to become comparable to those due to aspherical lunar gravity over longer time horizons. To directly compare these two perturbative sources, the ratio of their magnitudes is considered, using the log-scale to visualize their variations across locations on the grid: 
\begin{equation}
    \log_{10} \left( \frac{\overline{|\ddot{\mathbf{r}}|}_{\textrm{SRP}}}{\overline{|\ddot{\mathbf{r}}|}_{\textrm{Asph. Gravity}}}\right)
\end{equation}
Figure \ref{fig:log_rel_mag} shows the log-relative-magnitude over one period and one hundred periods (on the order of two to three weeks) for the locations on the grid.  

The log-relative-magnitude for the one period case ranges between -1.1 to -0.6 as shown in Fig. \ref{fig:comp_one_period}, indicating that the effects of aspherical lunar gravity dominate those of SRP across the considered locations. As the time horizon of interest is increase to one hundred orbital periods, however, the mean acceleration  magnitude for the aspherical lunar gravity reduce to the order of $10^{-4}$ m s$^{-2}$. This reduction may be explained by the collapse of various sectoral and tesseral terms of the spherical harmonics presented in Eq. \eqref{eq: spherical_harmonics} under first-order averaging techniques, with only the zonal terms and specific sectoral and tesseral terms meeting resonance conditions surviving over longer time horizons \cite{yarndley2026origins}. In comparison, the mean acceleration magnitudes due to SRP remain on the order of $10^{-4}$ m s$^{-2}$ over one hundred periods, with the effects being strongest at higher altitudes and near-polar inclinations where the mean sunlit fraction over the time horizon stays very close to 1. 

As shown in Fig. \ref{fig:comp_hundred_periods}, the log-relative-magnitude over one hundred periods tends closer to 0 for large portions of the domain, and is positive for locations with large altitudes and inclinations where the effects of SRP are dominant over those of aspherical lunar gravity. Given the dependence of the SRP on the physical characteristics of the fragments and the resulting spread of acceleration magnitudes, we primarily assess the time-evolution of non-sinking fragments under the first-order averaged effects of SRP, as derived in Section \ref{sec: milankovitch_evolution}. Note that this approach may be similarly applied to other perturbative sources, including through the use of closed-form averaged representations of various spherical harmonic terms in Eq. \eqref{eq: spherical_harmonics} developed as part of recent work by the European Space Agency (ESA) towards semi-analytical propagation of lunar RSOs \cite{efthymiopoulos2023selena, legnaro2024secular}. 

\subsection{Evolution of Non-Sinking Fragments Under SRP}\label{sec: milankovitch_evolution}
To consider the perturbative effects of SRP under a worst-case assumption of being sunlit at all times ($\nu = 1$), we utilize the Milankovitch orbital element vectors introduced in Section \ref{sec: milankovitch_background}. Recalling the definition of the SRP potential presented in Eq. \eqref{eq: srp_def} and denoting the Sun vector in the MCE frame as $\hat{\mathbf{r}}_\odot$, the averaged potential over one orbit of the RSO is written as: 
\begin{equation}
    \overline{\mathcal{R}}_{\textrm{SRP}} = \frac{1}{2\pi \sqrt{\mu a}} p_{\textrm{SRP}}\, C_\textrm{R}\, \psi \int_0^{2\pi} (\mathbf{r} \cdot \hat{\mathbf{r}}_{\odot}) \, dM,
\end{equation}
where $M$ is the mean anomaly of the RSO. Then, as $\omega_\odot \ll \dot{M}$: 
\begin{equation}
    \overline{\mathcal{R}}_{\textrm{SRP}} = \frac{1}{\sqrt{\mu a}}p_{\textrm{SRP}}\, C_\textrm{R}\, \psi \, \hat{\mathbf{r}}_{\odot} \cdot \left( \frac{1}{2\pi} \int_0^{2\pi} \mathbf{r} \, dM\right) =  \frac{1}{\sqrt{\mu a}} p_{\textrm{SRP}}\, C_\textrm{R}\, \psi \, (\overline{\mathbf{r}} \cdot\hat{\mathbf{r}}_{\odot}),
\end{equation}
with the value of $\overline{\mathbf{r}} =  -\frac{3}{2} a \mathbf{e}$ being a known result \cite{rosengren2013long}. Substituting this value: 
\begin{equation}\label{eq: renorm_R_srp}
    \overline{\mathcal{R}}_{\textrm{SRP}} = -\frac{3}{2} \sqrt{\frac{a}{\mu}} p_{\textrm{SRP}}\, C_\textrm{R} \, \psi \, (\mathbf{e} \cdot \hat{\mathbf{r}}_\odot) = \kappa_{\textrm{SRP}} \, \psi \, (\mathbf{e} \cdot \hat{\mathbf{r}}_\odot),
\end{equation}
where the consolidated leading coefficient $\kappa_{\textrm{SRP}} = -\frac{3}{2} \sqrt{\frac{a}{\mu}} p_{\textrm{SRP}}\, C_\textrm{R} $. The first-order averaged equations are thus obtained by substituting the derivatives of Eq. \eqref{eq: renorm_R_srp} into Eq. \eqref{eq: milankovitch_ode}, yielding: 
\begin{equation}
        \dot{\mathbf{h}} = \kappa_{\textrm{SRP}} \, \psi \, (\mathbf{e} \times \hat{\mathbf{r}}_\odot), \quad
        \dot{\mathbf{e}} = \kappa_{\textrm{SRP}} \, \psi \, (\mathbf{h} \times \hat{\mathbf{r}}_\odot)
\end{equation}

As the eccentricity vectors of initially non-sinking fragments evolve over time, fragments which sink over longer time horizons would need to have their eccentricity magnitudes sufficiently increase to cause their perilunes to decay. Since the secular drift in the semi-major axis for any fragment remains zero \cite{rosengren2013long, allan1964long}, the critical eccentricity value for a given fragment: 
\begin{equation}\label{eq: e_crit}
    e_{\textrm{critical}} = 1 - \frac{r_\textrm{M}}{a},
\end{equation}
is sufficient to identify such long time horizon sinking via the monitoring of $\| \mathbf{e}(t)\| \geq e_{\textrm{critical}}$ over time. In keeping with the formulation for short time horizon sinking fragments, the assessment of whether a fragment has attained this critical value need occur only at the beginning of each fragment orbital period to identify sinking. 

Thus, if for a fixed $\Delta t = T$ (the fragment orbital period, typically on the order of a few hours) the Sun vector is assumed to move sufficiently slowly, it may be held constant over that $\Delta t$ and denoted as $\hat{\mathbf{r}}_{\odot, k} = \hat{\mathbf{r}}_{\odot}(k\cdot\Delta t ; \lambda)$, with $\dot{\hat{\mathbf{r}}}_{\odot, k} = 0$. Then, the time-evolution of the orbital element vectors is represented by a set of difference equations for each period, obtained by computing the second derivatives from Eq. \eqref{eq: milankovitch_ode}: 
\begin{equation}\label{eq: milan_second_deriv}
    \begin{split}
        \ddot{\mathbf{e}} &= \kappa_{\textrm{SRP}} \, \psi (\hat{\mathbf{r}}_{\odot, k} \times \dot{\mathbf{h}}) = (\kappa_{\textrm{SRP}} \, \psi)^2 \, \hat{\mathbf{r}}_{\odot, k} \times (\hat{\mathbf{r}}_{\odot, k} \times \mathbf{e}) = (\kappa_{\textrm{SRP}} \, \psi)^2 \, [( \hat{\mathbf{r}}_{\odot, k}  \cdot \mathbf{e}) \hat{\mathbf{r}}_{\odot, k}  - \mathbf{e}]\\
        \ddot{\mathbf{h}} &= \kappa_{\textrm{SRP}} \, \psi (\hat{\mathbf{r}}_{\odot, k} \times \dot{\mathbf{e}}) = (\kappa_{\textrm{SRP}} \, \psi)^2 \, \hat{\mathbf{r}}_{\odot, k} \times (\hat{\mathbf{r}}_{\odot, k} \times \mathbf{h}) = (\kappa_{\textrm{SRP}} \, \psi)^2 \, [( \hat{\mathbf{r}}_{\odot, k}  \cdot \mathbf{h}) \hat{\mathbf{r}}_{\odot, k}  - \mathbf{h}]
    \end{split}
\end{equation}
As $\mathbf{\hat{r}}_{\odot, k} \perp (\mathbf{\hat{r}}_{\odot, k} \times \mathbf{e})$ and $\mathbf{\hat{r}}_{\odot, k} \perp (\mathbf{\hat{r}}_{\odot, k} \times \mathbf{h})$ and $(\hat{\mathbf{r}}_{\odot, k} \cdot \hat{\mathbf{r}}_{\odot, k}) = 1$, it can additionally be seen that: 
\begin{equation}
    \begin{split}
         \hat{\mathbf{r}}_{\odot, k} \cdot \dot{\mathbf{e}} &=  \kappa_{\textrm{SRP}} \, \psi [\hat{\mathbf{r}}_{\odot, k} \cdot(\hat{\mathbf{r}}_{\odot, k} \times \mathbf{h})] = 0\\
        \hat{\mathbf{r}}_{\odot, k} \cdot \ddot{\mathbf{e}} &= (\kappa_{\textrm{SRP}} \, \psi)^2 [-\hat{\mathbf{r}}_{\odot, k} \cdot \mathbf{e} + (\hat{\mathbf{r}}_{\odot, k}\cdot \mathbf{e})(\hat{\mathbf{r}}_{\odot, k} \cdot \hat{\mathbf{r}}_{\odot, k})] = 0\\
       \hat{\mathbf{r}}_{\odot, k} \cdot \dot{\mathbf{h}} &=  \kappa_{\textrm{SRP}} \, \psi [\hat{\mathbf{r}}_{\odot, k} \cdot(\hat{\mathbf{r}}_{\odot, k} \times \mathbf{e})] = 0 \\
       \hat{\mathbf{r}}_{\odot, k} \cdot \ddot{\mathbf{h}} &= (\kappa_{\textrm{SRP}} \, \psi)^2 [-\hat{\mathbf{r}}_{\odot, k} \cdot \mathbf{h} + (\hat{\mathbf{r}}_{\odot, k}\cdot \mathbf{h})(\hat{\mathbf{r}}_{\odot, k} \cdot \hat{\mathbf{r}}_{\odot, k})] = 0 \\
    \end{split}
\end{equation}
Both $\hat{\mathbf{r}}_{\odot, k} \cdot \mathbf{e}$ and $\hat{\mathbf{r}}_{\odot, k} \cdot \mathbf{h}$ are thus constant, and Eq. \eqref{eq: milan_second_deriv} for each vector simply represents a harmonic oscillator with a constant forcing term. The general solution to this system is: 
\begin{equation}\label{eq: milan_general_solution}
    \begin{split}
        \mathbf{e}(t) = A \cos(\kappa_{\textrm{SRP}}\, \psi \, t) + B \sin (\kappa_{\textrm{SRP}}\, \psi \, t) + (\hat{\mathbf{r}}_{\odot, k} \cdot \mathbf{e}) \hat{\mathbf{r}}_{\odot, k} \\
        \mathbf{h}(t) = C \cos(\kappa_{\textrm{SRP}}\, \psi \, t) + D \sin (\kappa_{\textrm{SRP}}\, \psi \, t) + (\hat{\mathbf{r}}_{\odot, k} \cdot \mathbf{h}) \hat{\mathbf{r}}_{\odot, k},
    \end{split}
\end{equation}
where $A, B, C, D$ are coefficients that can be determined using the boundary conditions for the given period i.e., $\mathbf{e}(0) = \mathbf{e}(k\cdot T) \equiv \mathbf{e}_k, \mathbf{h}(0) = \mathbf{h}_k$ and $\dot{\mathbf{e}}(0) = \dot{\mathbf{e}}_k, \dot{\mathbf{h}}(0) = \dot{\mathbf{h}}_k$. Substituting these conditions into Eq. \eqref{eq: milan_general_solution}, the coefficients are found to be: 
\begin{equation}
    \begin{split}
        A &= \mathbf{e}_k - (\hat{\mathbf{r}}_{\odot, k}\cdot \mathbf{e})\hat{\mathbf{r}}_{\odot, k}\\
        C &= \mathbf{h}_k - (\hat{\mathbf{r}}_{\odot, k}\cdot \mathbf{h})\hat{\mathbf{r}}_{\odot, k}
    \end{split}
    \quad 
    \begin{split}
        B &= \hat{\mathbf{r}}_{\odot, k}  \times \mathbf{h}_k\\
        D &= \hat{\mathbf{r}}_{\odot, k}  \times \mathbf{e}_k
    \end{split}
\end{equation}
Denoting $\kappa_{\textrm{SRP}} \, \psi \, T \equiv \xi$, the complete formulation of the per-period difference equations becomes: 
\begin{equation}\label{eq: milan_final_diff_eq}
    \begin{split}
        \mathbf{e}_{k+1} &=  \mathbf{e}_k \cos \xi + (\hat{\mathbf{r}}_{\odot, k} \times \mathbf{h}_k) \sin \xi + (\hat{\mathbf{r}}_{\odot, k} \cdot \mathbf{e}_k)\hat{\mathbf{r}}_{\odot, k} [1 - \cos \xi] \\
        \mathbf{h}_{k+1} &= \mathbf{h}_k \cos \xi + (\hat{\mathbf{r}}_{\odot, k} \times \mathbf{e}_k) \sin \xi + (\hat{\mathbf{r}}_{\odot, k} \cdot \mathbf{h}_k)\hat{\mathbf{r}}_{\odot, k}[1 - \cos \xi]
    \end{split}
\end{equation} 

The system of equations in Eq. \eqref{eq: milan_final_diff_eq} is linear, and the absolute value of the determinant of the Jacobian matrix for this system evaluates to 1 (as shown in Appendix \ref{sec: appendix_prob_det_milan}). Thus, for a probability density over $\mathbf{Y}_k = [\mathbf{h}_k, \mathbf{e}_k]$, the application of the difference equations as a change-of-variables becomes: 
\begin{equation}\label{eq:milan_density}
    \rho_{\mathbf{Y}_{k+1}} (\mathbf{Y}_{k+1}) = \frac{1}{|\det \mathbf{J}|} \rho_{\mathbf{Y}_k}(\mathbf{Y}_k) = \rho_{\mathbf{Y}_k}(\mathbf{Y}_k),
\end{equation}
which demonstrates that the time-evolution of the Milankovitch orbital element vectors in the MCE frame under the effects of SRP does not modify the volume of the probability mass of the distribution, and the critical eccentricity values described in Eq. \eqref{eq: e_crit} acts as a geometric constraint which captures the probability mass of fragments that sink within a given time horizon as $ \rho_{\mathbf{Y}_{k}}$ evolves over time. In order to define the initial distribution of these vectors, we note that the system has four degrees of freedom and can thus be recovered using a conditional distribution over the area-to-mass ratios for non-sinking fragments as well as the associated distribution over $\mathbf{\Delta v}$ values. 

\subsection{Initial Distribution Over the Milankovitch Vectors}\label{sec: nonsink_cond_amr}

To define an initial distribution over the Milankovitch orbital element vectors which may be used to assess the time-evolution of the non-sinking fragments, a conditional distribution over the area-to-mass ratio values $\psi$ for these fragments is first developed. Recalling the application of the push-forward measure in Eq. \eqref{eq:bern_integral} and the associated Bernoulli random variable $S$, the probability of not sinking within the first fragment orbital period may be denoted as $\mathbb{P}_S(S=0)$. Then, for the probability space over the area-to-mass ratios, denoted by the triple $(\Omega_{\psi}, \mathcal{F}_{\psi}, \mathbb{P}_{\psi})$ with $\Omega_\psi = \mathbb{R}_+$ and $\mathcal{F}_{\psi} = \mathcal{B}(\mathbb{R}_+)$, the conditional probability density over $\psi$ for the non-sinking fragments generated at a given originating location $(\mathbf{r}_O, \mathbf{v}_O)$ is: 
\begin{equation}
    \rho_{\psi | S } (\psi | S = 0) = \frac{\mathbb{P}_{S|\psi}(S = 0 | \psi) \cdot \rho_\psi (\psi)}{\mathbb{P}_S (S = 0)}  = \frac{\mathbb{P}_{S|\psi}(S = 0 | \psi) \cdot \rho_\psi (\psi)}{ 1 - p_s},
\end{equation}
where $p_s$ is the probability of sinking across for that originating location, as defined in Eq. \eqref{eq:bern_integral}, and $\rho_{\psi}$ is the probability density for $\mathbb{P}_{\psi}$. To obtain the likelihood $\mathbb{P}_{S | \psi}(S = 0 | \psi)$, the integral in Eq. \eqref{eq:bern_integral} is conditioned on $\psi$: 
\begin{equation}
    \mathbb{P}_{S|\psi}(S = 0 | \psi) = \int_{\Omega_{\text{non-sink}}} \rho_{\mathbf{\Delta v} | \psi}(\mathbf{\Delta v} | \psi) d \mathbf{\Delta v},
\end{equation}
where $\Omega_{\textrm{non-sink}} = \Omega^C_{\textrm{sink}}$ is the set of non-sinking fragments. Since the SBM internally uses the sampled value of $\psi$ for each fragment to sample the associated $|\mathbf{\Delta v}|$ value from a log-normal distribution, and the directional parameters $u$ and $\theta$ for the $\mathbf{\Delta v}$ are sampled independently, $\rho_{\mathbf{\Delta v} | \psi}$ is a known distribution. The four-dimensional input space over the non-sinking fragments from which the initial Milankovitch orbital element vectors are recovered is then constructed by recognizing that for each $\psi$ value sampled from $\rho_{\psi| S}(\psi | S = 0)$ the appropriate $\rho_{\mathbf{\Delta v} | \psi}$ can be identified, with the four-dimensional joint density being:
\begin{equation}\label{eq: milan_input_space}
    \rho_{(\psi, \mathbf{\Delta v}) | S}((\psi, \mathbf{\Delta v}) | S) = \rho_{\psi | S}(\psi | S = 0) \cdot \rho_{|\mathbf{\Delta v}| \mid \psi}(|\mathbf{\Delta v}| \mid \psi) \cdot \rho_u(u) \cdot \rho_\theta(\theta),
\end{equation}
where $\rho_u, \rho_\theta$ are the densities for the directional components of $\mathbf{\Delta v}$. Since the values of $|\mathbf{\Delta v}|$ and $\psi$ are joint by conditioning, a copula is used to connect the dependent marginal pair ($\psi, |\mathbf{\Delta v}|$) to the directional components ($u, \theta$) \cite{durante2010copula}. A copula $\mathcal{C}: [0, 1]^d \rightarrow [0, 1]$ is a mathematical structure linking the cumulative density functions (CDFs) $F_i$ of the univariate marginal distributions which make up a $d-$dimensional multivariate joint distribution with CDF $H$ as: 
\begin{equation}
    H(x_1, \dots , x_d) = \mathcal{C}(F_1(x_1), \dots, F_d(x_d))
\end{equation}
The probability density $h$ is similarly linked by the copula density $c$: 
\begin{equation}\label{eq: copula}
    h(x_1, \dots , x_d) = c(F_1(x_1), \dots, F_d(x_d)) \cdot f_1(x_1) \cdot \ldots \cdot f_d(x_d)
\end{equation}
In the case of the input space defined in Eq. \eqref{eq: milan_input_space}, the directional components are independent, and the marginals $\rho_{\psi | S}$ and $\rho_{|\mathbf{\Delta v}| \mid \psi}$ are used to fit a two-dimensional copula for the sampling of the four-dimensional density. Then, for a sample from this joint distribution, the $\mathbf{\Delta v}$ value is converted from the MCI to the MCE frame at the chosen epoch and used alongside the originating location $(\mathbf{r}_o, \mathbf{v}_o)$ to recover the corresponding Milankovitch vectors via Eq. \eqref{eq: milan_definition}. Since the input space is four dimensional, only the planar components of $\mathbf{h}$ and $\mathbf{e}$ as generated through this method are retained, and the out-of-plane components are found using the constraints defined in Eq. \eqref{eq: milan_constraints}. Then, denoting $\mathbf{Y}_0 = [\mathbf{h}(t_0), \mathbf{e}(t_0)]$ as a random vector of the initial Milankovitch vectors and $\rho_{\mathbf{Y}_0}$ the density of the same as the push-forward measure of Eq. \eqref{eq: milan_definition} on $\rho_{(\psi, \mathbf{\Delta v}) | S}$, the time-evolution of this sample using the difference equations in Eq. \eqref{eq: milan_final_diff_eq} can be carried out. Likewise, the associated $e_{\textrm{critical}}$ values for $\mathbf{Y}_0$ are calculated such that any additional sinking fragments over longer time horizons may be identified. 

\subsection{Fitting of Relevant Distributions}\label{sec: fitting_nonsink}

\begin{figure}[h!]
    \centering
    \begin{subfigure}[b]{0.45\textwidth}
         \centering
         \includegraphics[width=\textwidth]{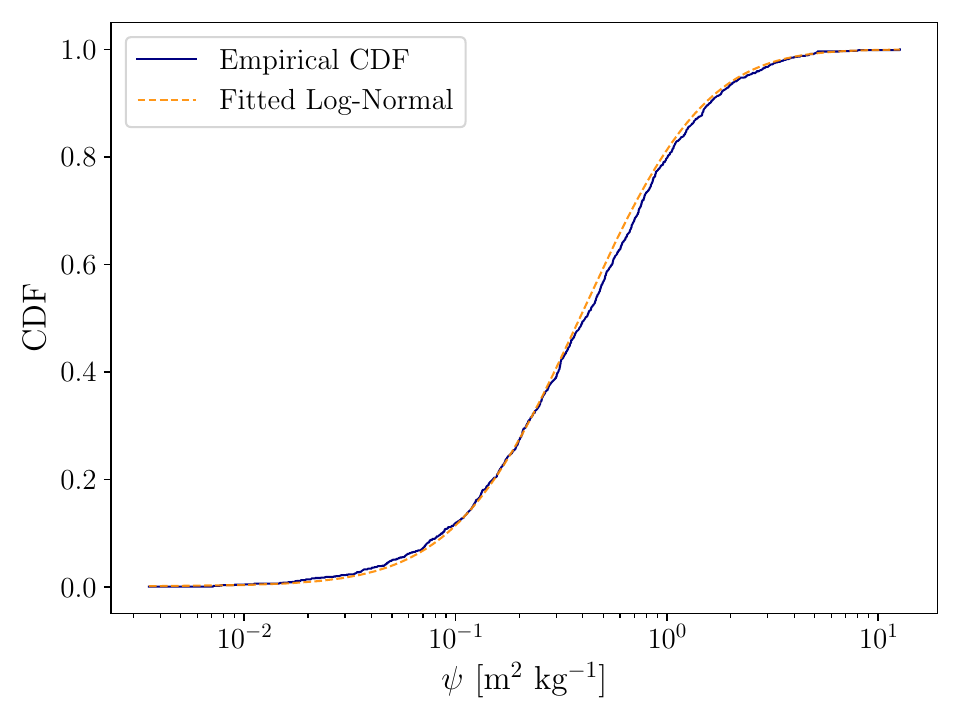}
         \caption{Fitted prior distribution over $\psi$, with $\mathbb{P}(D_n > d_n) = 0.463$ and $d_n = 0.022$.}
         \label{fig:fitted_psi_prior}
     \end{subfigure}
     \hfill
     \begin{subfigure}[b]{0.45\textwidth}
         \centering
         \includegraphics[width=\textwidth]{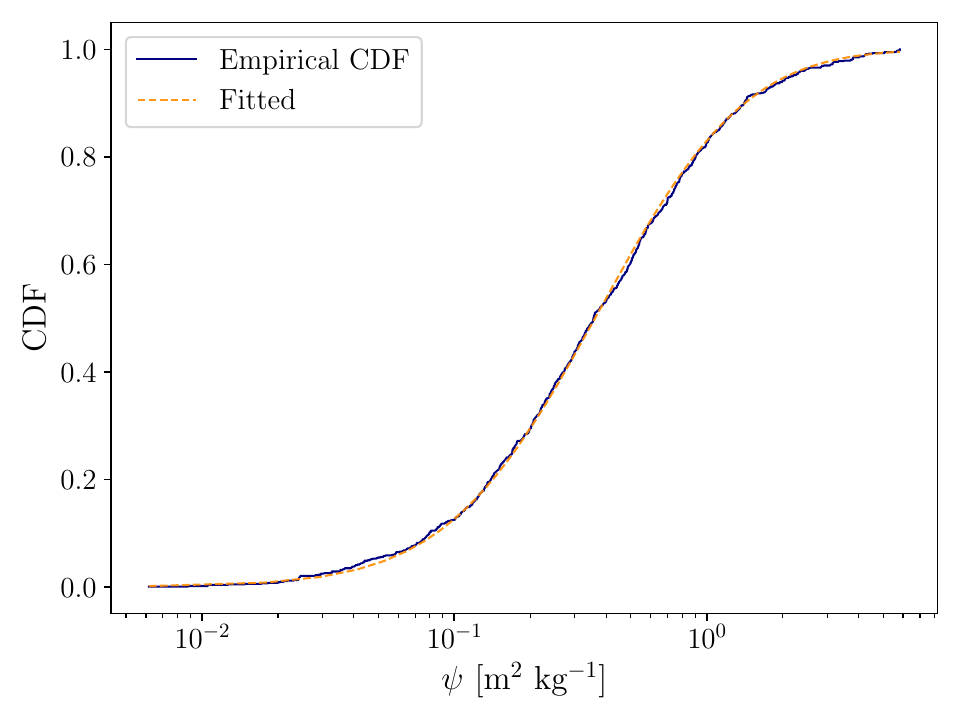}
         \caption{Calculated Posterior over $\psi | S = 0$, with $\mathbb{P}(D_n > d_n) = 0.703$ and $d_n = 0.020$.}
         \label{fig:posterior_psi_cond_ks}
     \end{subfigure}
     \vspace{1em}
     \begin{subfigure}[b]{0.45\textwidth}
         \centering
         \includegraphics[width=\textwidth]{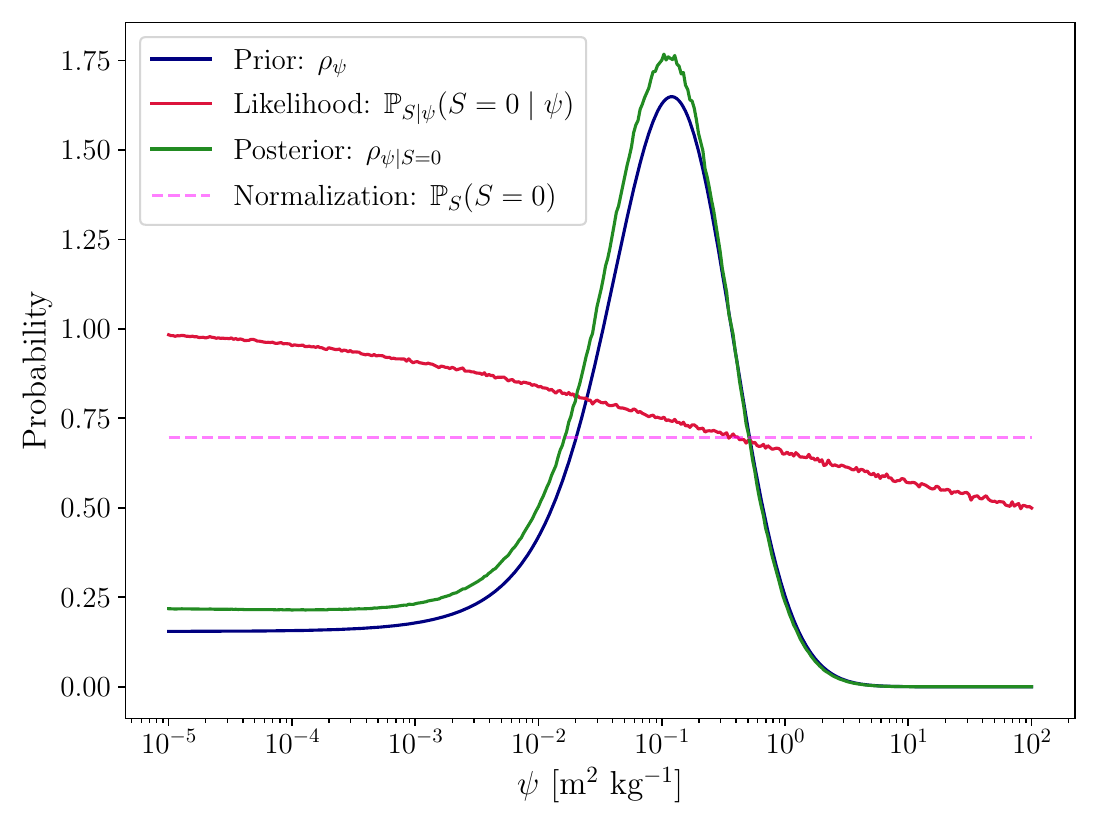}
         \caption{Components of posterior $\rho_{\psi|S}(\psi | S=0)$}
         \label{fig:posterior_psi_cond}
     \end{subfigure}
     \hfill
     \begin{subfigure}[b]{0.45\textwidth}
         \centering
         \includegraphics[width=\textwidth]{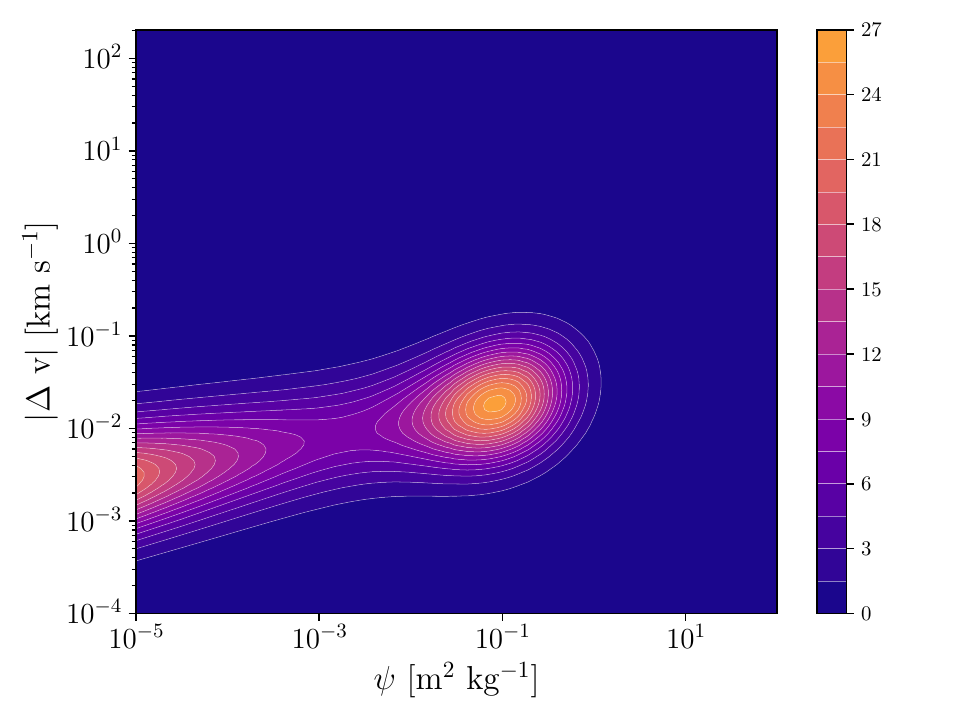}
         \caption{Joint density $\rho_{(\psi, |\mathbf{\Delta v}|)|S}((\psi, |\mathbf{\Delta v}|) | S=0)$}
         \label{fig:joint_prob_psi_dv}
     \end{subfigure}
    \caption{(top) Empirical and calculated CDFs of the prior (left) and posterior (right) distributions over $\psi$. (bottom) Posterior distribution over $\psi$ for the non-sinking fragments $(S = 0)$, with prior, likelihood, and normalization plotted (left), and the resulting posterior joint density via the fitted copula (right).}
    \label{fig:posterior_psi_overall}
\end{figure}

To fit a distribution over $\rho_{\psi | S}(\psi | S = 0)$ in order to recover the initial distribution over the Milankovitch orbital element vectors, the procedure outlined in Section \ref{sec: sinking_fitting} is followed, using the OpenTURNS package \cite{baudin2017open} in Python. A univariate distribution is fitted to the generated area-to-mass ratio values for sample of $2 \times 10^6$ fragments using the same model parameters as before, which serves as the prior distribution $\rho_\psi$. The best-fit distribution for $\rho_\psi$ is also found to be a log-normal distribution, with the parameters $\mu_{\log, \psi} = -0.91626$ and $\sigma_{\log, \psi} = 1.04487$, with Fig. \ref{fig:fitted_psi_prior} showing the result of the associated Kolmogorov-Smirnov one-sample test. 

The $p-$value $\mathbb{P}(D_n > d_n) = 0.463$ is well above the $\delta = 0.05$ threshold, and good visual agreement is seen between the empirical and fitted CDFs. We note that while the SBM itself uses two distributions for sampling $\psi$ based on the underlying $L_c$ value, alongside a bridging function, we may directly fit a single distribution using the samples generated by the SBM (which intrinsically capture the effects of $L_c$). While the full implementation of the set of SBM distributions may be necessary for modeling larger-scale breakup events (e.g., of the proposed Lunar Gateway or booster stages for delivering large RSOs) where a larger range of $L_c$ values is expected, a single fitted distribution for $\rho_\psi$ passes the Kolmogorov-Smirnov test with a sufficiently high $p-$value for the first-order averaging approach used by the proposed framework.

\begin{figure}[h!]
    \centering
    \includegraphics[width=0.95\linewidth]{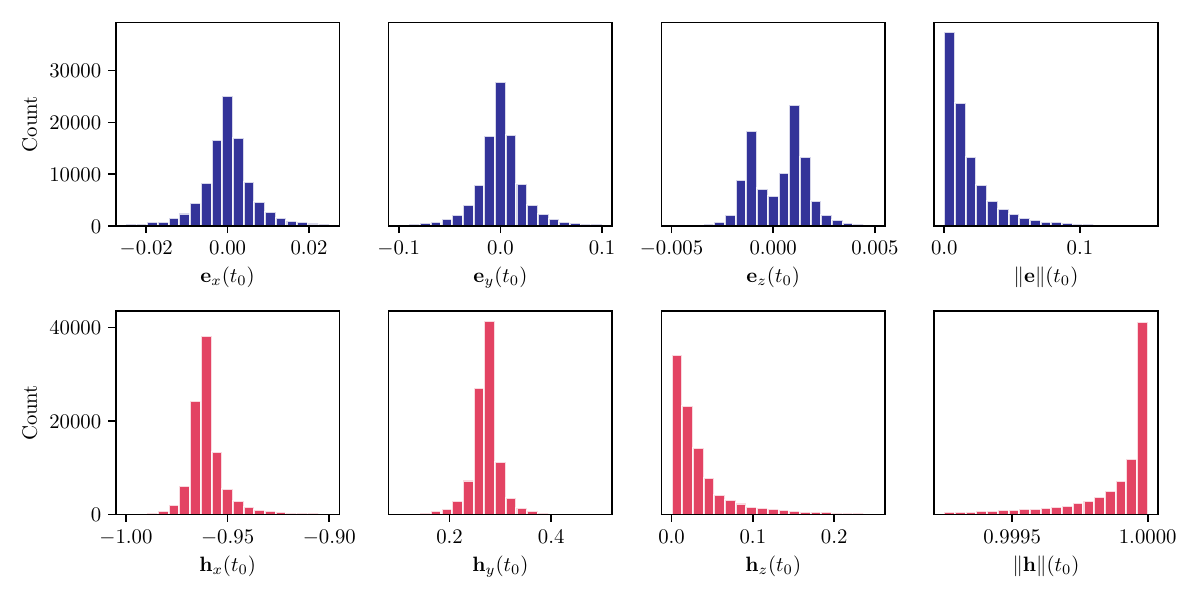}
    \caption{Histogram of initial Milankovitch orbital element vector components.}
    \label{fig:milan_init}
\end{figure}

The posterior distribution over the values of $\psi$ for the non-sinking fragments can then be calculated using the fitted prior distribution $\rho_\psi$, as described in Section \ref{sec: nonsink_cond_amr}. The resulting posterior density $\rho_{\psi | S}(S = 0)$ for an originating location $(\mathbf{r}_O, \mathbf{v}_O)$ on a 100 km circular polar LLO is shown in Fig. \ref{fig:posterior_psi_cond}. We can see that the likelihood decreases as $\psi$ increases, which is to be expected given that the magnitude of the associated $|\mathbf{\Delta v}|$ is higher for larger $\psi$ values and would be more likely to provide the sufficient change in velocity to $\mathbf{v}_O$ for the fragment to sink. Smaller $\psi$ values, on the other hand, lead to smaller changes in $\mathbf{v}_O$ which are more likely to slightly change the fragment's trajectory away from the originating orbit without sinking immediately. Fig. \ref{fig:posterior_psi_cond_ks} shows that the posterior distribution has a $p-$value $\mathbb{P}(D_n > d_n) = 0.703$ for a sample of size $\approx6750$ (using the non-sinking fragments from one debris generation event, with the total number of generated fragments being 9500 and the shown $\mathbb{P}(S = 0) \approx 0.7$), thus validating the goodness-of-fit of the posterior distribution. Using this distribution, a copula $\mathcal{C}$ is fit to obtain the joint density $\rho_{(\psi, |\mathbf{\Delta v}|)|S}((\psi, |\mathbf{\Delta v}|)|S=0)$ required to sample the four-dimensional input space, as shown in Fig. \ref{fig:joint_prob_psi_dv}. Figure \ref{fig:milan_init} shows an example of the components of $\mathbf{h}(t_0), \mathbf{e}(t_0)$ across $10^5$ samples, generated using $\rho_{(\psi, \mathbf{\Delta v}) | S}$ for the same originating location as in Fig. \ref{fig:posterior_psi_cond}. 

\subsection{Assessment of Non-Sinking Fragments}\label{sec: comparison_nonsinking}

\begin{figure}[h!]
    \centering
    \includegraphics[width=0.9\linewidth]{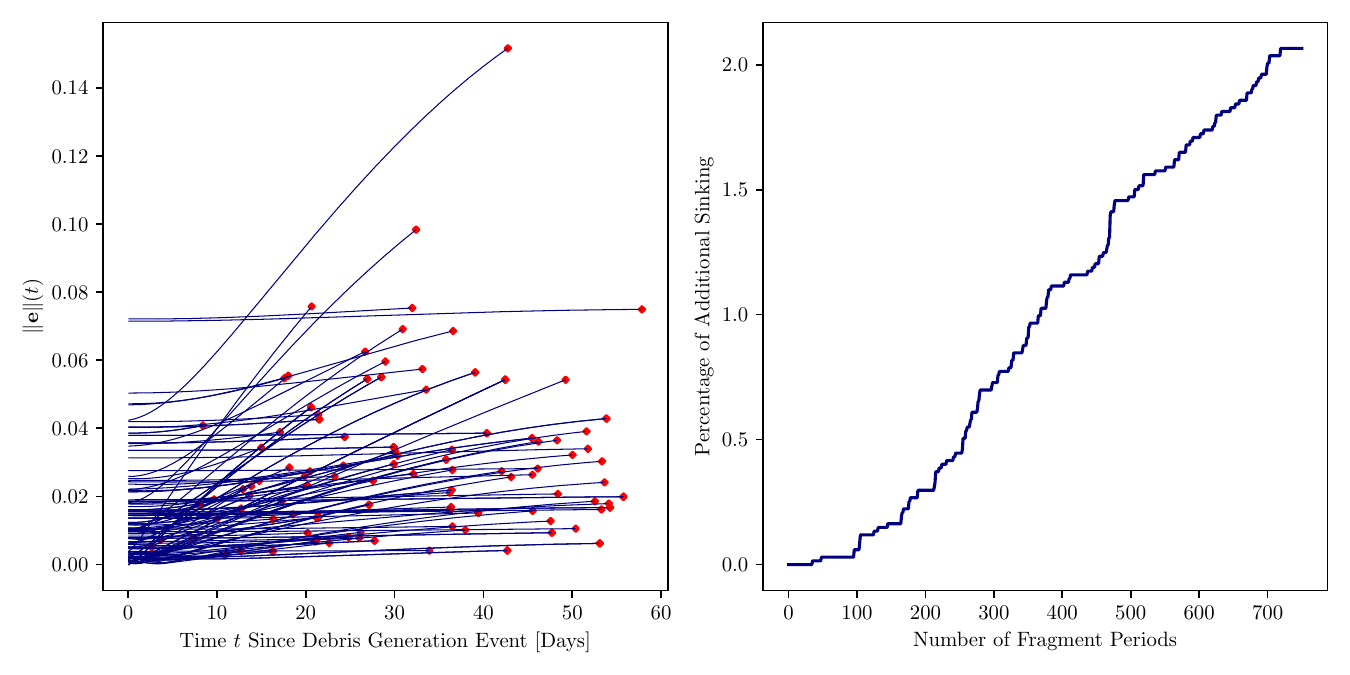}
    \caption{(left) Evolution of the eccentricity magnitude $\| \mathbf{e}\|(t)$ for samples which sink within 750 periods. (right) Percentage of additional sinking over 750 periods.}
    \label{fig:milan_main_result}
\end{figure}

Using the initial distribution of the Milankovitch orbital element vectors, $\rho_{\mathbf{Y}_0}$, the derived difference equations in Eq. \eqref{eq: milan_final_diff_eq} are applied to a sample of the initial states $\mathbf{Y}_0$ to observe whether any non-sinking fragments cross their $e_{\textrm{critical}}$ threshold values over longer time horizons. Figure \ref{fig:milan_main_result} shows the time-evolution of the eccentricity magnitude $\| \mathbf{e} \| (t)$ for samples generated at an originating location on a 100 km polar circular LLO which sink within 750 periods (ranging from 50-100 days depending on the fragment's orbital period). 

\begin{figure}[h!]
    \centering
    \begin{subfigure}[b]{0.45\textwidth}
         \centering
         \includegraphics[width=\textwidth]{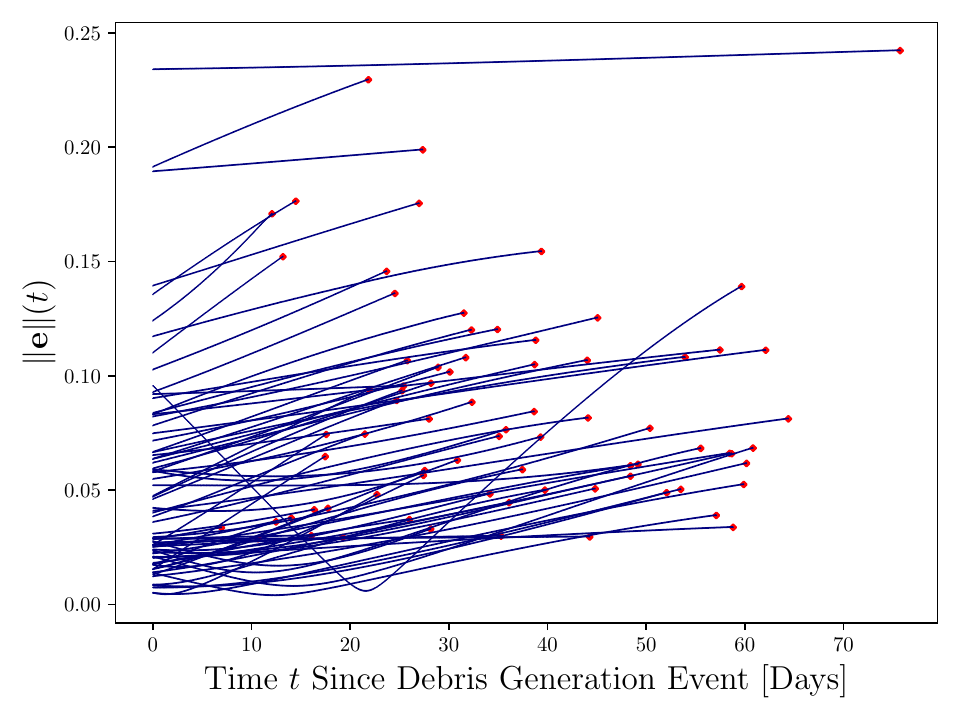}
         \caption{Monte Carlo validation of additional sinking.}
         \label{fig:nonsink_mc_valid}
     \end{subfigure}
     \hfill
     \begin{subfigure}[b]{0.42\textwidth}
         \centering
         \includegraphics[width=\textwidth]{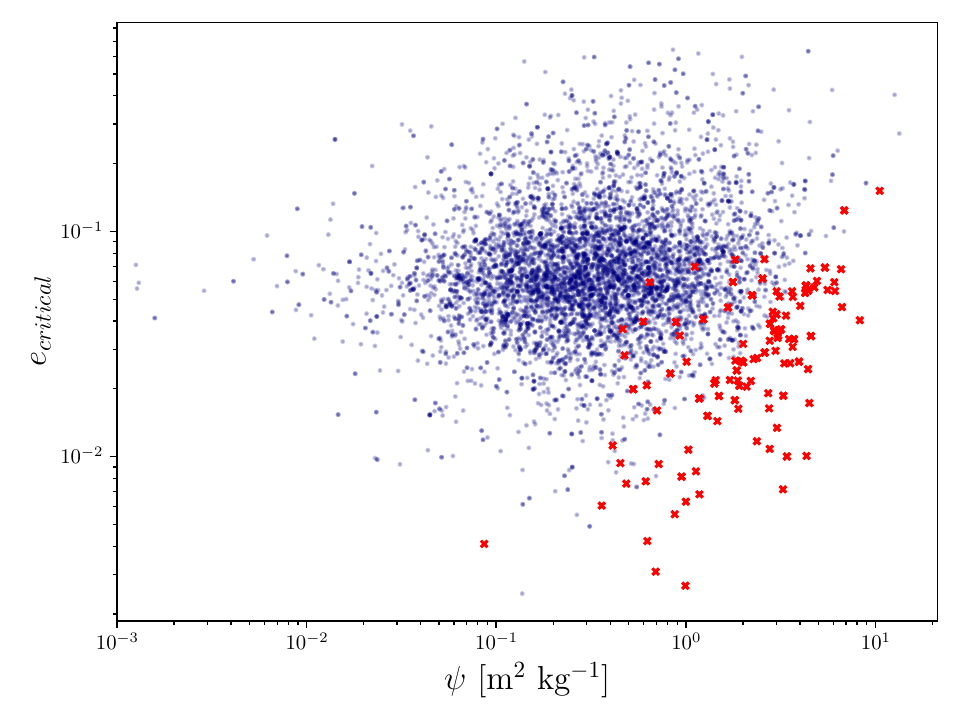}
         \caption{$\psi$ and $e_{\textrm{critical}}$ values for additional sinking (red) and lingering (navy) fragments.}
         \label{fig:psi_with_e_crit}
     \end{subfigure}
     \vspace{1 em}
     \begin{subfigure}[b]{0.85\textwidth}
         \centering
         \includegraphics[width=\textwidth]{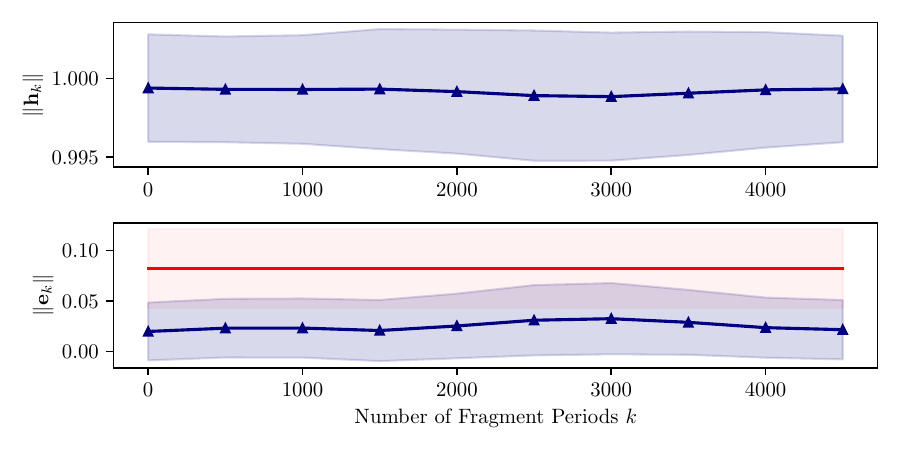}
         \caption{Evolution of $\| \mathbf{h}\|_k$ and $\| \mathbf{e}\|_k$ statistics (navy), with the $e_{\textrm{critical}}$ statistics (red) annotated.}
         \label{fig:milan_rot}
     \end{subfigure}
    \caption{Assessment of the time-evolution of the Milankovitch vectors for additional sinking fragments.}
    \label{fig:milan_valid_assess}
\end{figure}

The effects of SRP cause some fragments which did not sink within the first fragment orbital period to sink over longer time horizons, with up to 2\% of the total non-sinking population considered sinking within 750 fragment orbital periods (50 - 100 days). The time-evolution of eccentricity magnitudes for these additional sinking fragments shown in Fig. \ref{fig:milan_main_result} demonstrates the wide range of $e_{\textrm{critical}}$ values (denoted by the red markers) across the samples, with a correspondingly large range of eccentricity growth required to attain the same and sink to the lunar surface. For example, we see some fragments whose eccentricity magnitude grows slowly over time, with $e_{\textrm{critical}} - \| \mathbf{e} \| (t_0)$ on the order of $10^{-2}$; other fragments experience the effects of SRP more strongly, with $\| \mathbf{e}\|(t)$ growing by more than 0.1 until they sink. This additional sinking behavior is validated against a Monte Carlo propagation of fragments generated at the same originating location, as shown in Fig \ref{fig:nonsink_mc_valid}. 
The empirical behavior of the additional sinking fragments matches well with evolution of the Milankovitch eccentricity vector magnitudes, with similar percentages of additional sinking over 750 fragment orbital periods. Further, we can see from Fig. \ref{fig:psi_with_e_crit} that the additional sinking fragments within the considered time horizon tend to have higher area-to-mass ratio values and lower $e_{\textrm{critical}}$ values than the fragments which continue to linger in the LLO environment at the end of 750 fragment periods, as is expected due to the linear dependence of SRP on $\psi$. 

Lastly, we visualize the effects of the intersection of the critical eccentricity barrier with the the probability mass of $\rho_{\mathbf{Y}_k}$ as defined in Eq. \eqref{eq:milan_density} as it evolves over time, using the mean and standard deviation of $\| \mathbf{h} \|_k$ and $\| \mathbf{e} \|_k$ across all samples. Figure \ref{fig:milan_rot} shows the evolution of the $1\sigma$ probability mass of the samples over 5000 periods, with increases in the mean magnitude of $\mathbf{e}$ being mirrored by a corresponding decrease in the mean magnitude of $\mathbf{h}$ highlighting the preservation of the volume of the probability mass as well as the satisfaction of the constraints specified by Eq. \eqref{eq: milan_constraints}. The static geometric barrier formed by $e_{\textrm{critical}}$, visualized using the mean and standard deviation of the sample-specific values, intersects with the $1\sigma$ probability mass, indicating the regions where fragments may sink;  a larger intersecting region at any given time would lead to higher likelihood a given fragment having both the required $\| \mathbf{e} \|$ and $e_{\textrm{critical}}$ value to sink to the lunar surface.  Collectively, these results demonstrate the utility of the proposed framework to investigate the additional sinking of fragments over longer time horizons, identify the linkages between the physical characteristics of the fragments and their resulting outcomes, and study the time-evolution of their trajectories while avoiding the drawbacks of Monte Carlo techniques. 

\section{Use of Proposed Framework}\label{sec: results_utilization}

\begin{figure}[h!]
    \centering
    \includegraphics[width=\linewidth]{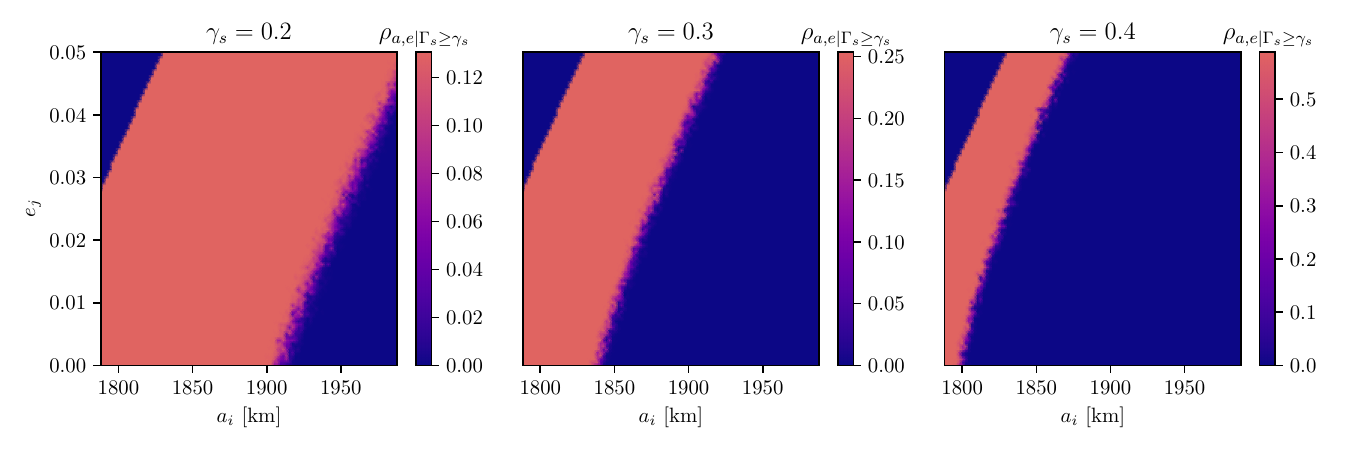}
    \caption{Posterior density $\rho_{a, e}(a_i, e_j | \Gamma_s \geq \gamma_s)$ for various values of $\gamma_s$, across the domain $\mathcal{D} = [1750 \textrm{ km}, 2050 \textrm{ km}] \times [0.0, 0.05]$.}
    \label{fig:policy_thresh}
\end{figure}

The utility of the probabilistic framework presented in this work has thus far been demonstrated as a standalone technique towards characterizing debris outcomes for a given originating location or debris generation event. The framework and its component distributions can additionally be developed and incorporated into broader-scale forecasting models, or applied towards the development of suitable policy thresholds for debris mitigation. To illustrate the same, we consider an example of policy development to mitigate long-term risks from lingering debris fragments. The probability of sinking in Eq. \eqref{eq: gamma} is defined for a fraction of exactly $\gamma_s$ of the total fragments sinking within one fragment orbital period; probabilistic thresholds for policy development may instead wish to consider the probability of the fraction of sinking fragments being  \textit{at least $\gamma_s$}, as represented by the binomial survival function:
\begin{equation}
\label{eq:binomial_survival}
    \mathbb{P}(\Gamma_s \geq \gamma_s) = \Sigma_{l = \lceil N\gamma_s \rceil}^N \binom{N}{l}p_s^{l} (1-p_s)^{N - l},
\end{equation}
which can be used to assess the compliance to some minimum sinking threshold $\gamma_s$ with probability $\mathbb{P}(\Gamma_s \geq \gamma_s) \geq p_{\textrm{thresh.}}$ for a given originating location. For mission designers to comply with this policy threshold, a useful extension of the proposed framework would entail the ability to rapidly survey the posterior probability density of originating locations and identify regions of the LLO domain where they may meet the required threshold probability $p_{\textrm{thresh.}}$. Using the representation of an originating location as the pair $(a_i, e_j)$ as defined in Section \ref{sec: sink_theory}, we may write for a known prior density $\rho_{a, e}$ the Bayesian posterior density over the originating locations given their probability of sinking is above $\gamma_s$ as: 
\begin{equation}\label{eq: bin_posterior}
\begin{split}
    \rho_{a, e}(a_i, e_j | \Gamma_s \geq \gamma_s) &= \frac{\rho_{a, e}(a_i, e_j) \cdot \mathbb{P}(\Gamma_s \geq \gamma_s | a = a_i, e = e_j)}{\mathbb{P}(\Gamma_s \geq \gamma_s)} \\[1em] 
    &= \frac{\rho_{a, e}(a_i, e_j) \cdot \mathbb{P}(\Gamma_s \geq \gamma_s | a = a_i, e = e_j)}{\iint_{\mathcal{D}} \mathbb{P}(\Gamma_s \geq \gamma_s | a = \tilde{a}, e = \tilde{e})\cdot \rho_{a, e}(\tilde{a}, \tilde{e}) \; d\tilde{a} \; d\tilde{e}},
\end{split}
\end{equation}
where $\mathcal{D} \subset \mathbb{R}_{+} \times [0, 1)$ is the domain over the $(a_i, e_j)$ pairs, and $\mathbb{P}(\Gamma_s \geq \gamma_s | a = a_i, e = e_j)$ is the evaluation of Eq. \eqref{eq:binomial_survival} using the sinking probability defined in Eq. \eqref{eq: sink_per_loc}. Figure \ref{fig:policy_thresh} shows the resulting `maps' of the posterior density defined in Eq. \eqref{eq: bin_posterior} for three values of $\gamma_s$. 
As the threshold $\gamma_s$ is increased, the region of viable locations in the orbital plane of interest reduces, and the probability density becomes more concentrated in these viable regions. Policymakers may thus use such tools to refine their proposed thresholds such that a balance may be struck between the viability of cislunar growth and the mitigation of risks to proximate missions in the immediate aftermath of a debris generation event. Mission designers may likewise use such maps for the given sinking threshold of interest to identify the orbital locations which are desirable from the perspective of mitigating debris risks, and then evaluate the specific probability $\mathbb{P}(\Gamma_s \geq \gamma_s)$ only at the locations of interest to demonstrate adherence to the policy threshold $p_{\textrm{thresh.}}$. In a similar manner, the theoretical development of the difference equations for the Milankovitch orbital element vectors may be extended to consider other perturbations of interest as well as to characterize the flux of debris fragments through the LLO domain, paving the way for the assessment of risks to proximate missions or incorporation into broader-scale models.

\section{Conclusion}\label{sec: conclusion}

This work presents the development of a probabilistic framework for the consideration of debris outcomes in low lunar orbits over short and medium time horizons, ranging from a few hours to a few months after a debris generation event. The first component of the proposed framework addresses outcomes over short time horizons and presents a formulation for the probability of debris fragments sinking within their first orbital period after the debris generation event. The fitting of relevant distributions is discussed, and the results of the formulation are shown to closely match the Monte Carlo baseline. For the fragments which do not sink within the first fragment orbital period, the effects of various perturbative sources are considered and the acceleration due to solar radiation pressure is identified as a dominant perturbation of interest. The Milankovitch orbital element vectors are used to define a first-order averaged time-evolution of the non-sinking fragments via a set of difference equations, and the additional sinking of these fragments over longer time horizons is assessed. The time-evolution of the probability mass over these vectors is shown to be volume-preserving, and the additional sinking behavior is demonstrated as the capturing of probability mass by a static geometric barrier formed by the critical eccentricity values for each fragment. Results from this approach are demonstrated to match the Monte Carlo propagation of trajectories, and linkages between the area-to-mass ratio values of the fragments and their resulting outcomes are additionally discussed. 

The proposed probabilistic framework may be refined through follow-on efforts to address limitations of scope and expand the scope of presented results. While both components of the framework are agnostic to the choice of parameters used for the SBM, several results have been demonstrated using the range of characteristic lengths from $0.01 $ m to $1$ m for the SBM as an intuitive baseline. If these parameters are changed, the distributions over $\mathbf{\Delta v}$ and $\psi$ would need to be fitted for samples from the SBM with the desired parameters, and the parameters of the fitted distributions would likewise change. Similarly, some results in this work have been presented using a circular polar LLO, given existing missions in similar orbits (e.g., the Lunar Reconnaissance Orbiter) and results for originating locations with vastly different orbital parameters, such as highly elliptical LLOs or orbits with very long periods, would need to relax any unsuitable assumptions when applying the framework. Additionally, future efforts may apply the first-order averaging approach to the remaining perturbative sources as well, strengthening the assessment of additional sinking over longer time horizons. 

Collectively, the proposed probabilistic framework provides a mechanism for the assessment of the probability of sinking and non-sinking outcomes over these transient time horizons, providing a novel contribution that complements the existing literature on the steady-state debris outcomes for various LLOs. The presented framework and its components may additionally be extended or adapted to for the development of probabilistic policy thresholds aimed at mitigating debris risks, and can likewise inform the compliance to these thresholds for mission designers during preliminary orbit selection. Future work towards the development of this framework may consider extensions to model the flux of debris from a given debris generation event through the LLO environment over time, leveraging the time-evolution of the Milankovitch orbital element vectors to map the evolving flux density. The theoretical basis of both components of the framework may also be adapted to other orbital families within the cislunar domain, eventually enabling the characterization of debris flux between families and the development of associated inter-family risk metrics. 

\section*{Funding Sources}
\noindent This material is based upon work supported by the Air Force Office of Scientific Research under Grant \#FA9550-25-1-0100.

\section*{Acknowledgments}

\noindent Simulations were performed on computational resources managed and supported by Princeton Research Computing, a consortium of groups including the Princeton Institute for Computational Science and Engineering (PICSciE) and the Office of Information Technology’s High-Performance Computing Center and Visualization Laboratory at Princeton University. The authors would additionally like to acknowledge Catherine Spence, an undergraduate student at Princeton University, for her efforts and discussions towards exploratory Monte Carlo analyses that informed the development of this work and Jannik Graebner, a PhD candidate at Princeton University, for his feedback on this manuscript. 

\newpage

\bibliography{sample}

\newpage 
\appendix
\setcounter{section}{0}
\renewcommand\thesection{\Alph{section}} 

\section{Determinant of the Milankovitch Vector Difference Equations}\label{sec: appendix_prob_det_milan}

To derive the determinant of the difference equations governing the time-evolution of the Milankovitch orbital element vectors presented in Eq. \eqref{eq: milan_final_diff_eq}, we first rewrite these equations as the application of a linear operator. We define a $\mathbb{R}^{3\times3}$ matrix for the cross-product: 
\begin{equation}
\renewcommand{\arraystretch}{0.6}
    [\hat{\mathbf{r}}_{\odot, k}]_\times = \begin{bmatrix}
        0 & 0 & \sin(\omega_\odot \cdot kT + \lambda)\\
        0 & 0 & -\cos(\omega_\odot \cdot kT + \lambda)\\
        -\sin(\omega_\odot \cdot kT + \lambda)& \cos(\omega_\odot \cdot kT + \lambda)& 0
    \end{bmatrix}
\end{equation}
where $T$ is the fragment period and $\lambda$ is the phase shift from the J2000 epoch, as previously defined. Likewise, for the outer product: 
\begin{equation}
\renewcommand{\arraystretch}{0.6}
    [\hat{\mathbf{r}}_{\odot, k} \hat{\mathbf{r}}_{\odot, k}^T] \equiv [\hat{\mathbf{r}}_{\odot, k}]_\otimes = \begin{bmatrix}
        \cos^2(\omega_\odot \cdot kT + \lambda)& \cos(\omega_\odot \cdot kT + \lambda)\sin(\omega_\odot \cdot kT + \lambda)& 0 \\
        \cos (\omega_\odot \cdot kT + \lambda) \sin(\omega_\odot \cdot kT + \lambda)& \sin^2({\omega_\odot \cdot kT + \lambda})& 0 \\
        0 & 0 & 0
    \end{bmatrix}
\end{equation}
Then, denoting $\xi = \kappa_{SRP}\, \psi\, T$ we have: 
\begin{equation}
\renewcommand{\arraystretch}{0.6}
    \begin{bmatrix}
        \mathbf{h}_{k+1} \\ \mathbf{e}_{k+1}
    \end{bmatrix} = \begin{bmatrix}
        \cos(\xi)\mathbf{I}_3 + (1-\cos(\xi))[\hat{\mathbf{r}}_{\odot, k}]_\otimes& \sin(\xi)[\hat{\mathbf{r}}_{\odot, k}]_\times\\
        \sin(\xi)[\hat{\mathbf{r}}_{\odot, k}]_\times & \cos(\xi)\mathbf{I}_3 + (1-\cos(\xi))[\hat{\mathbf{r}}_{\odot, k}]_\otimes
    \end{bmatrix}\begin{bmatrix}
        \mathbf{h}_{k} \\ \mathbf{e}_{k}
    \end{bmatrix}
\end{equation}
where the linear operator is the Jacobian, $\mathbf{J}$ of the system. Note now that $[\hat{\mathbf{r}}_{\odot, k}]_\otimes[\hat{\mathbf{r}}_{\odot, k}]_\times = [\hat{\mathbf{r}}_{\odot, k}]_\times[\hat{\mathbf{r}}_{\odot, k}]_\otimes = \mathbf{0}_3$, which makes the two sub-blocks on the diagonal equivalent (we can call them $\mathbf{J}_1$), and on the off-diagonal also equivalent ($\mathbf{J}_2$), with $\mathbf{J}_1\mathbf{J}_2 = \mathbf{J}_2\mathbf{J}_1$. Also note that $([\hat{\mathbf{r}}_{\odot, k}]_\times)^2 = -\mathbf{I}_3 + [\hat{\mathbf{r}_{\odot, k}}]_\otimes$, as can be shown by expanding the left-hand side. Thus, the determinant of the full Jacobian matrix then becomes: 
\begin{equation}
\renewcommand{\arraystretch}{0.6}
    \begin{split}
        \det \mathbf{J} &= \begin{vmatrix}
            \mathbf{J}_1 & \mathbf{J}_2 \\ \mathbf{J}_2 & \mathbf{J}_1
        \end{vmatrix} = \det(\mathbf{J}_1-\mathbf{J}_2)\det(\mathbf{J}_1+\mathbf{J}_2) = \det(\mathbf{J}^2_1 - \mathbf{J}_2^2) \\
        \mathbf{J}_1^2 &= \cos^2(\xi)\mathbf{I}_3 + 2\cos(\xi)\mathbf{I}_3(1-\cos(\xi))[\hat{\mathbf{r}}_{\odot, k}]_\otimes + (1-\cos(\xi))^2\underbrace{[\hat{\mathbf{r}}_{\odot, k}]_\otimes^2}_{= [\hat{\mathbf{r}}_{\odot, k}]_\otimes} \\
        &= \cos^2(\xi)\mathbf{I}_3 + (1 - \cos{(\xi)})(1+\cos(\xi))[\hat{\mathbf{r}}_{\odot, k}]_\otimes = \cos^2(\xi)\mathbf{I}_3 + \sin^2{(\xi)}[\hat{\mathbf{r}}_{\odot, k}]_\otimes\\
        \mathbf{J}_2^2 &= (\sin(\xi)[\hat{\mathbf{r}}_{\odot, k}]_\times)^2 = \sin^2(\xi) \left[- \mathbf{I}_3 + [\hat{\mathbf{r}}_{\odot, k}]_{\otimes}\right] \\ 
        \mathbf{J}^2_1 - \mathbf{J}^2_2 &=  \cos^2(\xi)\mathbf{I}_3 + \sin^2{(\xi)}[\hat{\mathbf{r}}_{\odot, k}]_\otimes - \sin^2(\xi)[\hat{\mathbf{r}}_{\odot, k}]_{\otimes} + \sin^2{(\xi)}\mathbf{I}_3 = \mathbf{I}_3 \\
        \implies& \det \mathbf{J} = 1
    \end{split}
\end{equation}

\end{document}